\documentclass[
 reprint,
 amsmath,amssymb,
 pra,
 floatfix,
]{revtex4-2}

\usepackage{graphicx}
\usepackage{dcolumn}
\usepackage{bm}
\usepackage{hyperref}
\hypersetup{colorlinks = true, 
  urlcolor = blue, linkcolor = blue, citecolor = blue}

\begin{document}

\preprint{APS/123-QED}

\title{Time evolution of coherent wave propagation and spin relaxation\\in spin-orbit coupled systems}
\author{Masataka Kakoi}
\email{kakoi@presto.phys.sci.osaka-u.ac.jp}
\author{Keith Slevin}
\email{slevin.keith.sci@osaka-u.ac.jp}
\affiliation{Department of Physics, Osaka University, 1-1 Machikaneyama-cho, Toyonaka, Osaka 560-0043, Japan}
\date{\today}
\begin{abstract}
    We investigate, both numerically and analytically, the time evolution of a particle in an initial plane wave 
    state as it is subject to elastic scattering in a two-dimensional disordered system with Rashba spin-orbit coupling (SOC).
    In the analytic calculation, we treat the SOC non-perturbatively, and the disorder perturbatively using the Diffuson and the Cooperon.
    We calculate the time dependence of 
    coherent backscattering (CBS) as a function of the strength of the SOC.
    We identify weak and strong SOC regimes, and give the relevant time and energy scales in each case.
    By studying the time dependence of the anisotropy of the disorder-averaged momentum distribution we identify the spin relaxation time.
    We find a crossover from D'yakonov-Perel' spin relaxation for weak SOC to Elliot-Yafet like behaviour for strong SOC.
\end{abstract}

\maketitle

\section{Introduction\label{sec:introduction}}

Cold atomic gases have unique features that are advantageous for research on coherent wave propagation 
in disordered systems.
Experiments can be performed in a regime where the gas is non-interacting. 
The gas can be put into a specified initial state and the subsequent time evolution of the spatial and 
momentum distributions of the atoms can be measured. 
The gas can be subjected to a random potential by applying laser speckle.
Observations~\cite{Jendrzejewski_2012_PRL,Labeyrie_2012} of CBS,
and observations~\cite{Billy_2008,*Roati_note,*Roati_2008,Kondov_2011, Jendrzejewski_2012_Nat_Phys} of Anderson localisation,
have been performed in this way.
Further developments of experimental technique e.g., bichromatic speckle~\cite{Lecoutre_2022} promise quantitative
measurement of the Anderson transition in speckle potentials.

An alternative approach is to subject the gas to a quasi-periodic modulation to realise
a quantum kicked rotor with a quasi-periodic kick.
This system exhibits dynamical localisation, an analogue of Anderson localisation in momentum space.
In this way, the analogue of the three-dimensional (3D) Anderson transition has been observed~\cite{Chabe_2008}.
The critical exponent of the transition has been measured~\cite{Lopez_2012} and found to be in good agreement
with numerical finite time scaling studies of the quantum kicked rotor~\cite{Lemarie_2009} and
with finite size scaling studies of Anderson’s model of localisation~\cite{Slevin_1999, Slevin_2014}.
Other signatures of Anderson localisation, namely coherent forward scattering and the quantum boomerang effect
have been proposed theoretically~\cite{Karpiuk_2012,Prat_2019} and then
subsequently observed~\cite{Hainaut_2018,Sajjad_2022} in the cold atom quantum kicked rotor.

CBS results from interference between time reversed scattering processes.
This interference is usually constructive and is manifest in an
enhanced probability, ideally by a factor two, for a wave to be scattered in the direction opposite 
to that of the incident wave.
This has been observed in optics~\cite{Kuga_1984,VanAlbada_1985,*Wolf_1985}, acoustics~\cite{Bayer_1993}, and in cold atoms~\cite{Jendrzejewski_2012_PRL,Labeyrie_2012}.
In solid state physics, it is manifest in the weak localisation effect~\cite{Bergmann_1984_review}.
In cold atoms, the constructive interference of time-reversed states 
leads to the emergence of characteristic structures in the disorder-averaged momentum distribution~\cite{Cherroret_2012_PRA, Cherroret_2021_review}.  

SOC is well known in atomic and solid state physics. 
It involves the coupling of an electron's spin with its orbital motion.
The importance of SOC is that it breaks spin rotation symmetry while preserving
time reversal symmetry.
In a system with both time reversal symmetry and spin rotation symmetry, the electron spin plays no role in the dynamics and the operation of time reversal corresponds to reversing 
the electron's momentum.
In a system with time reversal symmetry but where spin rotation symmetry is broken by SOC, the operation of time reversal also involves reversing the electron's spin.
This has a dramatic effect on the interference between time reversed processes, changing it from constructive to destructive.
In solid state physics this is manifested in the weak anti-localisation effect~\cite{Hikami_1980, Bergmann_1984_review}.
Synthetic SOC has been realised experimentally in both Bose and Fermi gases 
\cite{Lin_2011, Wang-P_2012, *Cheuk_2012, Huang_2016, Wu_2016, Wang-ZY_2021}. 
This brings into prospect experiments that combine synthetic SOC with random potentials
and this has stimulated recent theoretical work~\cite{Yue_2020, Wang-Q_2021, Janarek_2022, Alluf_2023, Arabahmadi_2024, Janarek_preprint}.

\begin{figure*}
  \includegraphics[width=0.95\linewidth]{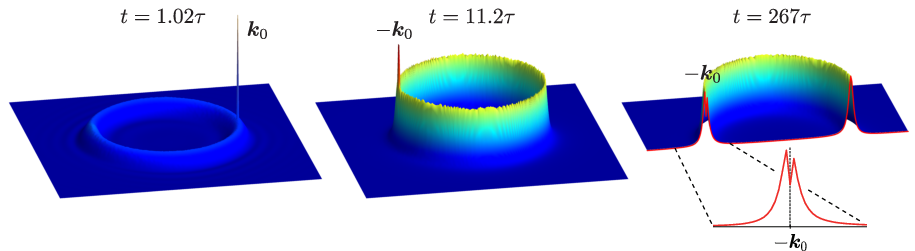}
  \caption{\label{fig:3d_plot_weak}
  (Color online) Simulation of the time evolution of a plane wave with initial wave vector $\bm{k}_0=(k_0,0)$ 
  in 2D systems with weak disorder and weak SOC. 
  We use the Ando model~\cite{Ando_1989}, as described by Eq.~(\ref{eq:Hamiltonian_Ando_real_spin_basis}), with
  $\varphi=\pi/1024$, and a disorder strength of $W=1$. 
  The disorder-averaged momentum distribution $n(\bm{k},t)$ is estimated by sampling $2^{13}=8192$ disorder realisations. 
  An isotropic diffusive background ring due to elastic scattering is observed with a crossover from enhanced backscattering at a time of several $\tau$ to reduced backscattering at much longer times.
  }
\end{figure*}

To make clear the importance of SOC for interference between time reversed scattering processes, 
consider a gas of electrons in the independent particle approximation 
described by a single particle Hamiltonian $H$.
Let us suppose that $H$ has time reversal symmetry, i.e., that $H$ commutes with a time reversal operator $T$
\begin{equation} \label{eq:time reversal symmetry}
   \left[ H , T \right] = 0.
\end{equation}
If $H$ has spin rotation symmetry, we have $T^2=+1$ since reversing the momentum twice leaves the momentum unchanged. 
If spin rotation symmetry is broken by SOC, we have $T^2=-1$. 
This is because reversing the electron's spin twice introduces a change of sign.
Starting from Eq.~(\ref{eq:time reversal symmetry}) and assuming $T^2=-1$, we find that the probability of scattering 
into a time reversed state is zero (see Appendix~\ref{app:matrix_element})
\begin{equation}\label{eq:no-scattering_to_time_reversed_state}
   \left\langle T \psi_0 \right| \exp \left(-i H t\right) \left| \psi_0 \right\rangle = 0,
\end{equation}
for any arbitrary initial state $\left|\psi_0\right\rangle$ and for all times $t$.
Note that this does not mean that scattering into a state with opposite momentum,
cannot occur. Rather that scattering into a state with opposite momentum and opposite spin cannot occur.

\begin{table}[b]
    \caption{
    The energy scales and the time scales governing the time-dependent behaviour for the weak and strong SOC limits.
    Here, $\tau$ is the scattering time, and $\Delta$ is the spin splitting induced by SOC. 
    \label{tab:time_scale}}
    \begin{ruledtabular}
    \renewcommand{\arraystretch}{1.5}
    \begin{tabular}{ccc}
     & weak SOC & strong SOC\\
    \hline
    energy scales & $\Delta\ll\hbar/\tau$ & $\Delta\gg\hbar/\tau$\\
    time scales & $\tau$, $\hbar^2/(\tau\Delta^2)$ & $\tau$, $\hbar/\Delta$ \\
    \end{tabular}
    \end{ruledtabular}
\end{table}

\begin{figure*}
  \includegraphics[width=0.94\linewidth]{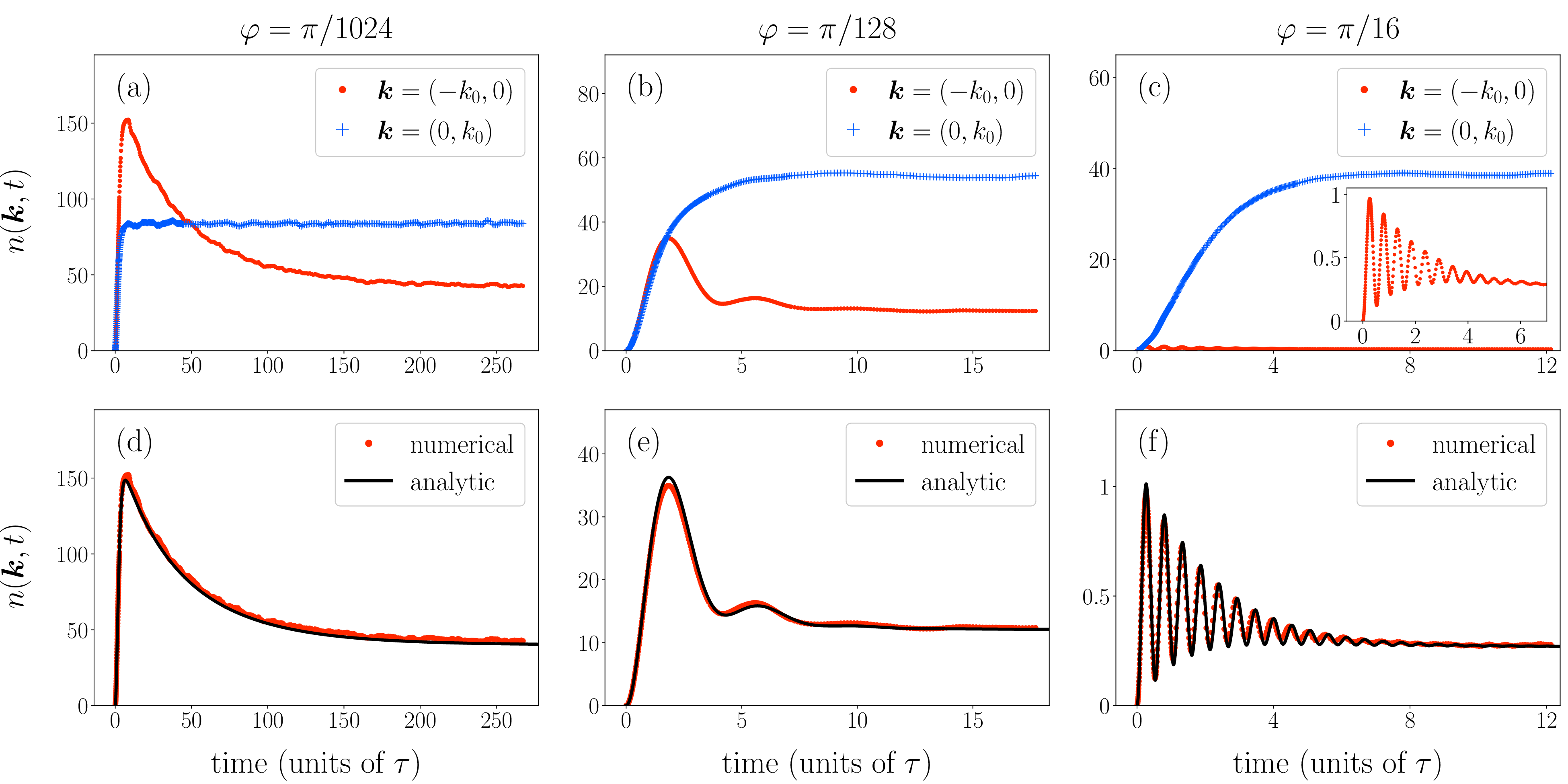}
  \caption{
  (Color online) (a)-(c)~Comparison of the time dependence of the disorder-averaged momentum distribution $n(\bm{k},t)$ at two points $\bm{k}=(-k_0,0)$~(red points) and $\bm{k}=(0,k_0)$~(blue crosses) estimated by sampling $8192$ disorder realisations. 
  The disorder strength is fixed at $W=1$ for all cases, while the strength of the SOC varies: $\varphi=\pi/1024$ for weak SOC, $\varphi=\pi/128$ for intermediate SOC, and $\varphi=\pi/16$ for strong SOC. 
  (d)-(f)~Comparison between the disorder-averaged momentum distribution at $\bm{k}=(-k_0,0)$ obtained in the simulation and that
  obtained from a diagrammatic expansion~(solid line).
  Equation~(\ref{eq:analytic_n_CBS_weak_SOC}) is plotted in (d), while Eq.~(\ref{eq:analytic_n_CBS_strong_SOC}) is plotted in (e) and (f).
  }
  \label{fig:time_dependence_CBS}
\end{figure*}

\begin{figure}[tbp]
  \includegraphics[width=0.7\linewidth]{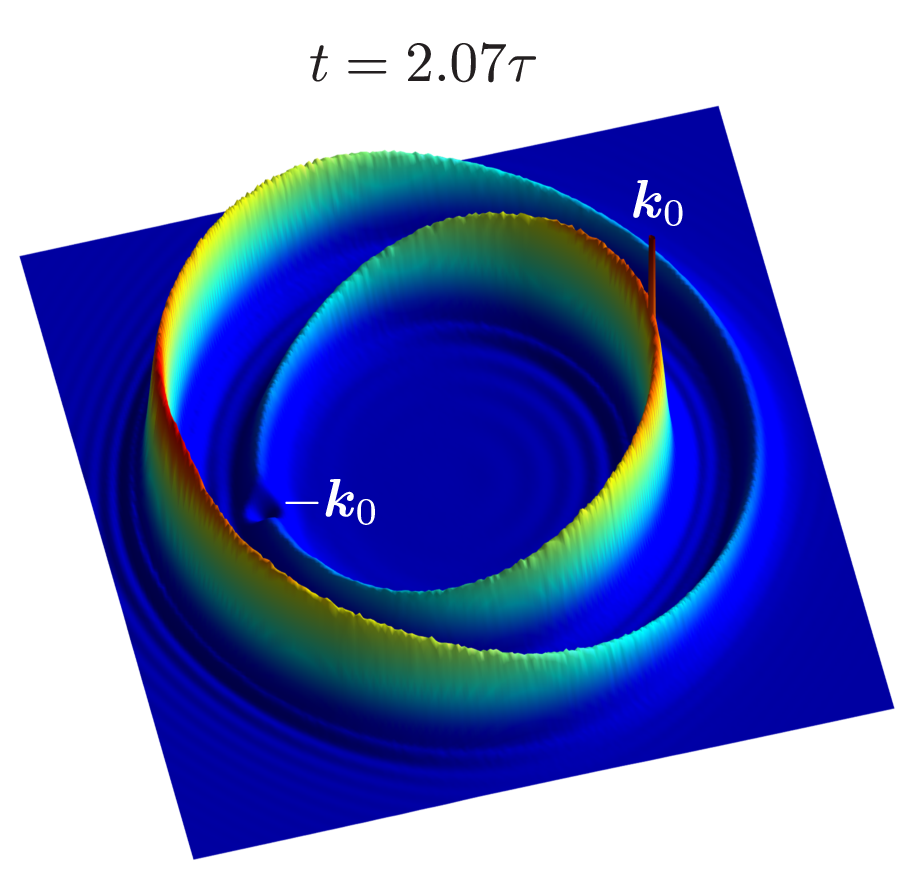}
  \caption{\label{fig:3d_plot_strong}
  (Color online) Disorder-averaged momentum distribution for strong SOC. 
  The model parameters are $\varphi=\pi/16$ and $W=1$. 
  At long times, the anisotropy of the diffusive background rings disappears.
  A view of the cross-section is presented in the top of
  Fig.~\ref{fig:anisotropic_background_along_xy-axis}.
  }
\end{figure}

In this paper, we consider the time evolution of an initial plane wave state with wave vector $\bm{k}_0$
resulting from elastic scattering by the random potential.
By treating the SOC non-perturbatively, and disorder perturbatively using a diagrammatic perturbation theory,
we identify weak and strong SOC regimes. In Table~\ref{tab:time_scale} we tabulate the corresponding
time and energy scales.
In Fig.~\ref{fig:3d_plot_weak}, in a regime of weak SOC, we show the build up of the diffusive ring 
on the order of
the scattering time $\tau$, the enhancement of backscattering on times of the order of several $\tau$,
and the reduction of backscattering at much longer times.
As can be seen in Fig.~\ref{fig:time_dependence_CBS}(d)-(f), 
we obtain good agreement between numerical and analytical results without any fitting parameters
for the entire range from weak SOC to strong SOC. 

SOC also leads to significant anisotropy in the disorder-averaged momentum distribution.
This is seen not only for strong SOC, as in Figs.~\ref{fig:3d_plot_strong},~\ref{fig:anisotropic_background} and~\ref{fig:anisotropic_background_along_xy-axis} but also for weak SOC when
the disorder-averaged momentum distribution is spin resolved as in Fig.~\ref{fig:hidden-anisotropy_weak_SOC}.
This anisotropy relaxes at sufficiently long times due to spin relaxation.
By studying this, we are able to determine the spin relaxation time and investigate the crossover
from D'yakonov-Perel' spin relaxation for weak SOC to Elliot-Yafet like behaviour for strong SOC.

\section{Model\label{sec:model}}

In this study, we use the Ando model~\cite{Ando_1989}, which is a tight-binding model with Rashba-type SOC defined on a square lattice, with the lattice constant taken as the unit of length. 
The Hamiltonian in the spin basis
\begin{equation}
    |\uparrow\rangle = \left(
    \begin{array}{c}
         1  \\
         0 
    \end{array}
    \right),\ \ \ 
    |\downarrow\rangle = \left(
    \begin{array}{c}
         0 \\
         1 
    \end{array}
    \right),
\end{equation}
is given by
\begin{multline}\label{eq:Hamiltonian_Ando_real_spin_basis}
    H_0 = -t_{\mathrm{hop}}\sum_{\bm{r}}\sum_{\sigma,\sigma'}\left[\left(\mathrm{T}_x\right)_{\sigma\sigma'}c^{\dag}_{\bm{r}+\bm{\mathrm{e}}_x,\sigma}c^{}_{\bm{r},\sigma'}\right.\\
    \left.+\left(\mathrm{T}_y\right)_{\sigma\sigma'}c^{\dag}_{\bm{r}+\bm{\mathrm{e}}_y,\sigma}c^{}_{\bm{r},\sigma'}+h.c.\right].
\end{multline}
Here,
\begin{equation}
    \mathrm{T}_x = \left(
    \begin{array}{cc}
        t_1 & t_2\\
        - t_2 & t_1
    \end{array}
    \right),\ 
    \mathrm{T}_y = \left(
    \begin{array}{cc}
        t_1 & - i t_2\\
        - i t_2 & t_1
    \end{array}
    \right),
\end{equation}
with
\begin{equation}
        t_1 = \cos\varphi,\ t_2=\sin\varphi. 
\end{equation}
Also, $c^{\dag}_{\bm{r}, \sigma}$ and $c^{}_{\bm{r}, \sigma}$ are creation and annihilation operators at site $\bm{r}$ and spin $\sigma$, and $\bm{\mathrm{e}}_x$ and $\bm{\mathrm{e}}_y$ are lattice vectors.
We take $t_{\mathrm{hop}}$ and $\hbar/t_{\mathrm{hop}}$ as the units of energy and time, respectively.

We consider square systems with linear size $L$.
The Hamiltonian~(\ref{eq:Hamiltonian_Ando_real_spin_basis}) may then be expressed in momentum space as
\begin{equation}
    H_0 = \sum_{\sigma,\sigma'}\sum_{\bm{k}} \big(\mathrm{H}_0(\bm{k})\big)_{\sigma\sigma'}\,c^{\dag}_{\bm{k},\sigma}c^{}_{\bm{k},\sigma'},
\end{equation}
where the summation is over spin and the appropriate allowed values of $\bm{k}$ in the 1st Brillouin zone, and
\begin{align}\label{eq:Hamiltonian_Ando_momentum_spin_basis}
    \nonumber
    &\mathrm{H}_0(\bm{k}) \\
    &=\left(
    \begin{array}{cc}
        -2t_1(\cos k_x+\cos k_y) & 2t_2(-i\sin k_x - \sin k_y)\\
        2t_2(i\sin k_x - \sin k_y) & -2t_1(\cos k_x+\cos k_y)
    \end{array}
    \right).
\end{align}

\begin{figure}[tbp]
  \includegraphics[width=0.8\linewidth]{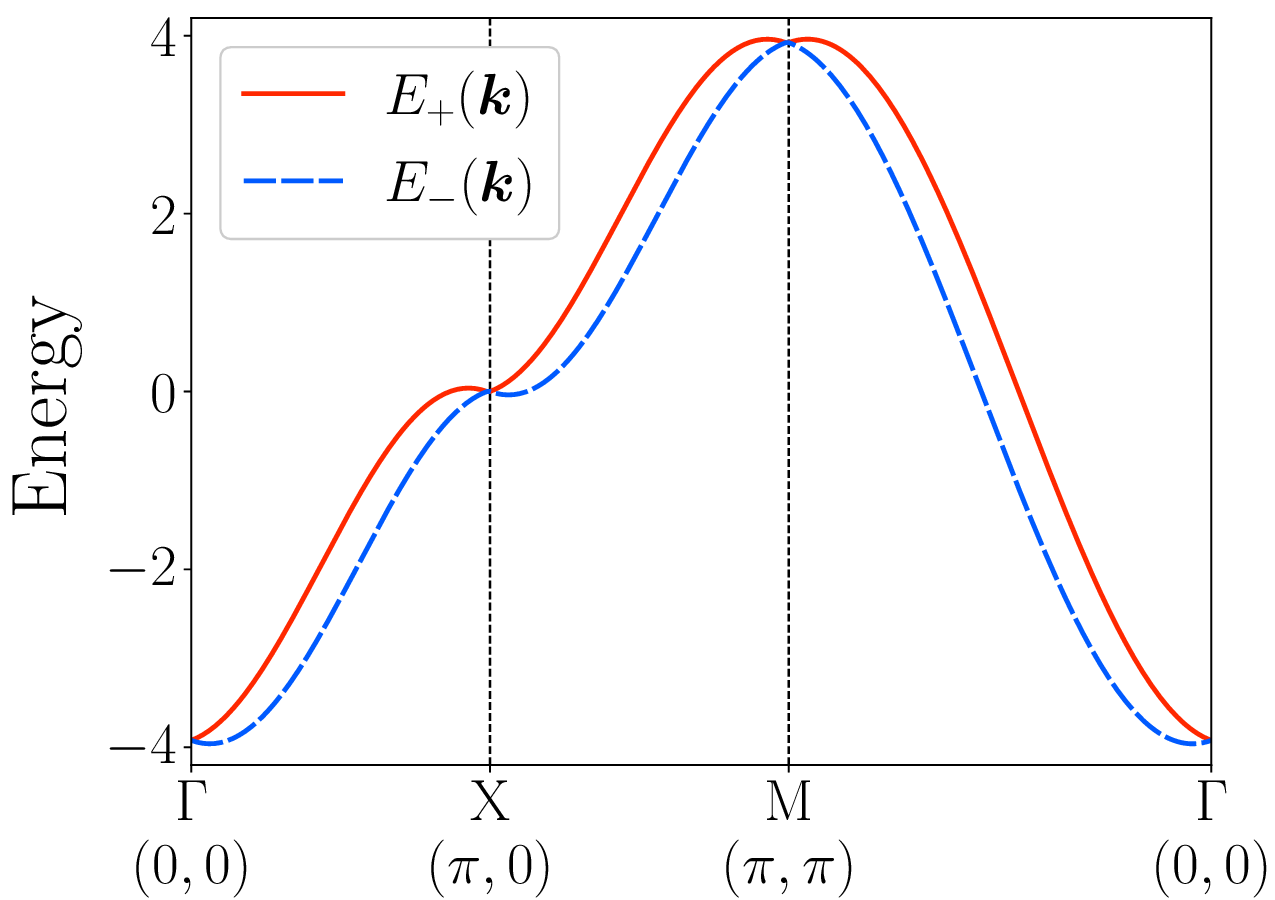}
  \caption{\label{fig:Ando_band}
  (Color online) Band structure of Ando model with parameter $t_1 = \cos\varphi$, $t_2 = \sin\varphi$, $\varphi=\pi/16$. As $t_2$ increases, the spin splitting into upper branch $E_+(\bm{k})$ and lower branch $E_-(\bm{k})$ becomes larger.}
\end{figure}

The Hamiltonian~(\ref{eq:Hamiltonian_Ando_momentum_spin_basis}) is diagonalised by the following momentum-coupled spin basis, which we refer to in what follows as the $\pm$~basis,
\begin{align}
    \label{eq:band_plus_state}
    |\bm{k},+\rangle &= |\bm{k}\rangle\otimes\left(\frac{|\uparrow\rangle - e^{-i\theta(\bm{k})}|\downarrow\rangle}{\sqrt{2}}\right),\\
    \label{eq:band_minus_state}
    |\bm{k},-\rangle &= |\bm{k}\rangle\otimes\left(-\frac{e^{i\theta(\bm{k})}|\uparrow\rangle +|\downarrow\rangle}{\sqrt{2}}\right),
\end{align}
where $\theta(\bm{k})$  is a real value for any wave vector $\bm{k}$ and satisfies the following, 
\begin{equation}\label{eq:def_theta}
    e^{i\theta(\bm{k})} = \frac{i\sin k_x + \sin k_y}{\sqrt{\sin^2k_x + \sin^2 k_y}}.
\end{equation}
The state $|-\bm{k},+\rangle$ is equal to the time-reversed state of $|\bm{k},+\rangle$ to within a phase factor~\cite{Time-reversal_note},
\begin{align}\label{eq:TR-state1}
    \nonumber
    T|\bm{k},+\rangle &= i\sigma_y K |\bm{k},+\rangle\\
    \nonumber
    &= |-\bm{k}\rangle\otimes\left(-\frac{|\downarrow\rangle + e^{i\theta(\bm{k})}|\uparrow\rangle}{\sqrt{2}}\right)\\
    &= e^{i\theta(\bm{k})}|-\bm{k},+\rangle,
\end{align}
where $\sigma_y$ is the Pauli matrix and $K$ the complex conjugation, and we used the relation $e^{i\theta(-\bm{k})} = -e^{i\theta(\bm{k})}$.
Also, the corresponding relationship holds for $|-\bm{k},-\rangle$ and $|\bm{k},-\rangle$,
\begin{equation}\label{eq:TR-state2}
    T|\bm{k},-\rangle = e^{-i\theta(\bm{k})}|-\bm{k},-\rangle.
\end{equation}
The eigenstates~(\ref{eq:band_plus_state}) and (\ref{eq:band_minus_state}) have eigenenergies
\begin{align}\label{eq:eigenenergy_Ando}
    \nonumber
    E_{\pm}(\bm{k}) &= -2t_1(\cos k_x + \cos k_y) \\
    &\quad\pm 2t_2\sqrt{\sin^2k_x + \sin^2k_y},
\end{align}
respectively.
In Fig.~\ref{fig:Ando_band}, we show the band structure of the Ando model, where the spin splitting into upper and lower branches due to SOC is visible. 

We suppose a spin independent random potential of the form
\begin{equation}\label{eq:random_potential}
    V =  \sum_{\bm{r}}\sum_{\sigma}w_{\bm{r}}\,c^{\dag}_{\bm{r}, \sigma}c^{}_{\bm{r}, \sigma},
\end{equation}
where $w_{\bm{r}}$ are independently and identically
distributed random variables with uniform distribution on the interval $[-W/2,W/2]$.
The potential distribution is uncorrelated and translationally invariant.
Therefore, the disorder-averaged Green's function is also translationally invariant, i.e.,
\begin{equation}
  \overline{G}^{R/A}(\bm{r},\bm{r}',\varepsilon) = \overline{G}^{R/A}(\bm{r}-\bm{r}',\varepsilon),
\end{equation}
and is diagonal with respect to the wave vector~\cite{Akkermans_2007_book},
\begin{equation}
    \overline{G}^{R/A}(\bm{k},\bm{k}',\varepsilon) = \overline{G}^{R/A}(\bm{k},\varepsilon)\,\delta_{\bm{k},\bm{k}'}.
\end{equation}
Here, $R$ stands for "retarded" and $A$ stands for "advanced". 
In addition, if the random potential is spin independent and has short-range correlations, as in this paper, the disorder-averaged Green's function is a diagonal matrix in the $\pm$~basis,
\begin{align}\label{eq:disorder-averaged_Greens_func_band_basis}
    \nonumber
    \overline{G}^{R/A}(\bm{k},\varepsilon) &= g_+^{R/A}(\bm{k},\varepsilon)\,|\bm{k},+\rangle\langle\bm{k},+|\\
    &\quad +g_-^{R/A}(\bm{k},\varepsilon)\,|\bm{k},-\rangle\langle\bm{k},-|.
\end{align}
The diagonal elements are
\begin{align}
    \label{eq:disorder-averaged_retarded_Greens_func_band_basis}
    g_{\pm}^R(\bm{k}, \varepsilon) &= \frac{1}{\varepsilon-E_{\pm}(\bm{k})+i/2\tau},\\
    \label{eq:disorder-averaged_advanced_Greens_func_band_basis}
    g_{\pm}^A(\bm{k}, \varepsilon) &= \frac{1}{\varepsilon-E_{\pm}(\bm{k})-i/2\tau}.
\end{align}
The detailed calculation is presented in Appendix~\ref{app:Green's func}.

\vspace{5pt}
\section{Diagrammatic Calculation\label{sec:diagrammatic_calculation}}

In this study, we consider the time evolution of the disorder-averaged momentum distribution. 
The disorder-averaged momentum distribution at wave vector $\bm{k}$ and time $t$, given an initial state $|\psi_0\rangle$, is 
\begin{align}\label{eq:def_n}
    \nonumber
    n(\bm{k},t) &= \overline{\langle\psi(t)|\Big(|\bm{k},+\rangle\langle\bm{k},+| + |\bm{k},-\rangle\langle\bm{k},-|\Big)|\psi(t)\rangle}\\
    &:= n_+(\bm{k},t) + n_-(\bm{k},t),
\end{align}
where
\begin{equation}
    |\psi(t)\rangle = e^{-i(H_0+V)t}|\psi_0\rangle.
\end{equation}
The branch resolved components $n_+(\bm{k})$ and $n_-(\bm{k})$ can be expressed as~\cite{Cherroret_2012_PRA, Kuhn_2005,*Kuhn_2007, Ghosh_2014, Scoquart_2020, integral_note}
\begin{equation}\label{eq:n_exact}
    n_{\pm}(\bm{k},t) = \int_{-\infty}^{\infty}\frac{d\varepsilon}{2\pi}\int_{-\infty}^{\infty}\frac{d\omega}{2\pi}e^{-i\omega t}\mathit{\Phi}_{\pm}(\bm{k},\varepsilon,\omega),
\end{equation}
where
\begin{widetext}
\begin{equation}\label{eq:def_Phi}
    \mathit{\Phi}_{\pm}(\bm{k},\varepsilon,\omega) = \sum_{a,b=\pm}\int\frac{d^2\bm{k}'\, d^2\bm{k}''}{(2\pi)^4}\,
    \overline{\left\langle\bm{k},\pm\middle|G^R\left(\varepsilon+\frac{\omega}{2}\right)\middle|\bm{k}',a\right\rangle\left\langle\bm{k}'',b\middle|G^A\left(\varepsilon-\frac{\omega}{2}\right)\middle|\bm{k},\pm\right\rangle}\left\langle\bm{k}',a|\psi_0\right\rangle\left\langle\psi_0|\bm{k}'',b\right\rangle.
\end{equation}
To avoid confusion, from now on, we associate Latin subscripts with $\pm$ basis and Greek subscripts with spin basis i.e., $a$ and $b$ represent $+$ or $-$, while $\alpha$, $\beta$ and so on will represent $\uparrow$ or $\downarrow$.
As the initial state, we take the plane wave that is an eigenstate of $H_0$ given in Eq.~(\ref{eq:Hamiltonian_Ando_real_spin_basis}) in upper branch whose energy is $E_+(\bm{k}_0)$ 
\begin{equation}\label{eq:initial_state}
    |\psi_0\rangle = |\bm{k}_0, +\rangle.
\end{equation}
Substituting the initial state~(\ref{eq:initial_state}) into Eq.~(\ref{eq:def_Phi}), performing the integration on $\bm{k}'$ and $\bm{k}''$ and summing over $a$ and $b$, we obtain
\begin{equation}\label{eq:Phi_all_scattering_process}
    \mathit{\Phi}_{\pm}(\bm{k},\varepsilon,\omega) = \overline{\left\langle\bm{k},\pm\middle|G^R\left(\varepsilon+\frac{\omega}{2}\right)\middle|\bm{k}_0,+\right\rangle\left\langle\bm{k}_0,+\middle|G^A\left(\varepsilon-\frac{\omega}{2}\right)\middle|\bm{k},\pm\right\rangle}.
\end{equation}
Equation~(\ref{eq:Phi_all_scattering_process}) includes contributions from all scattering processes, but we approximate these by retaining only the contributions from the ladder diagram (Diffuson) and the maximally crossed diagram (Cooperon)~\cite{Akkermans_2007_book, Cherroret_2012_PRA},
\begin{equation}\label{eq:n_initial_state_k0_+}
    n_{\pm}(\bm{k},t) \simeq \int_{-\infty}^{\infty}\frac{d\varepsilon}{2\pi}\int_{-\infty}^{\infty}\frac{d\omega}{2\pi}e^{-i\omega t}\left[\mathit{\Phi}^0_{\pm}(\bm{k},\varepsilon,\omega) + \mathit{\Phi}^D_{\pm}(\bm{k},\varepsilon,\omega) + \mathit{\Phi}^C_{\pm}(\bm{k},\varepsilon,\omega)\right],
\end{equation}
where
\begin{align}
    \displaybreak[0]
    \mathit{\Phi}^0_{\pm}(\bm{k},\varepsilon,\omega) &= g_+^R\left(\bm{k}_0,\varepsilon+\frac{\omega}{2}\right)g_+^A\left(\bm{k}_0,\varepsilon-\frac{\omega}{2}\right)\delta(\bm{k}-\bm{k}_0)\,\delta_{\pm,+},\\
    \displaybreak[0]
    \label{eq:Phi^D}
    \mathit{\Phi}^D_{\pm}(\bm{k},\varepsilon,\omega) &=  g_{\pm}^R\left(\bm{k},\varepsilon+\frac{\omega}{2}\right)g_{\pm}^A\left(\bm{k},\varepsilon-\frac{\omega}{2}\right)\mathit{\Gamma}^D_{\pm\pm,++}(\bm{k},\bm{k}_0, \varepsilon,\omega)\,g_+^R\left(\bm{k}_0,\varepsilon+\frac{\omega}{2}\right)g_+^A\left(\bm{k}_0,\varepsilon-\frac{\omega}{2}\right),\\
    \label{eq:Phi^C}
    \mathit{\Phi}^C_{\pm}(\bm{k},\varepsilon,\omega) &=  g_{\pm}^R\left(\bm{k},\varepsilon+\frac{\omega}{2}\right)g_{\pm}^A\left(\bm{k},\varepsilon-\frac{\omega}{2}\right)\mathit{\Gamma}^C_{\pm\pm,++}(\bm{k},\bm{k}_0, \varepsilon,\omega)\,g_+^R\left(\bm{k}_0,\varepsilon+\frac{\omega}{2}\right)g_+^A\left(\bm{k}_0,\varepsilon-\frac{\omega}{2}\right).
\end{align}
\end{widetext}
The operators $\mathit{\Gamma}^D$ and $\mathit{\Gamma}^C$ in the $\pm$~basis can be projected to the operators $\Gamma^D$ and $\Gamma^C$ in the spin basis using unitary operators with
\begin{equation}
    U_{\alpha\beta,ab}(\bm{k}) = \langle\bm{k},\alpha|\bm{k},a\rangle \langle\bm{k},b|\bm{k},\beta\rangle,
\end{equation}
as elements. The projection is represented in matrix representation as follows,
\begin{align}
    \label{eq:transform_Gamma^D_spin_band}
    \mathit{\Gamma}^{D}(\bm{k},\bm{k}',\varepsilon,\omega) &= U^{\dag}(\bm{k})\Gamma^{D}(\bm{0},\varepsilon,\omega)U(\bm{k}'),\\
    \label{eq:transform_Gamma^C_spin_band}
    \mathit{\Gamma}^{C}(\bm{k},\bm{k}',\varepsilon,\omega) &= U^{\dag}(\bm{k})\Gamma^{C}(\bm{k}+\bm{k}',\varepsilon,\omega)U(\bm{k}'),
\end{align}
where $U(\bm{k})$ is a Hermitian unitary matrix,
\begin{equation}
    U(\bm{k}) = \frac12\left(
    \begin{array}{cccc}
        1 & -e^{i\theta(\bm{k})} & -e^{-i\theta(\bm{k})} & 1 \\
        -e^{-i\theta(\bm{k})} & -1 & e^{-2i\theta(\bm{k})} & e^{-i\theta(\bm{k})} \\
        -e^{i\theta(\bm{k})} & e^{2i\theta(\bm{k})} & -1 & e^{i\theta(\bm{k})} \\
        1 & e^{i\theta(\bm{k})} & e^{-i\theta(\bm{k})} & 1
    \end{array}
    \right).
\end{equation}
The diagrams of $\Gamma^D$ and $\Gamma^C$ are shown in Fig.~\ref{fig:diagram}.
In the absence of SOC, time-reversal symmetry ensures that reversing the direction of the arrows yields the same result. 
Consequently, in the long-time limit, the contributions of the Diffuson and the Cooperon become equal at $\bm{k}=-\bm{k}_0$. 
However, in the present discussion, the direction of the arrows carries significant meaning.
For $\Gamma^D$ and $\Gamma^C$ in the spin basis, Bethe-Salpeter equations
\begin{align}
    \label{eq:Bethe-Salpeter_Eq_Diffuson}
    \nonumber
    \displaybreak[0]
    &\Gamma^D_{\alpha\beta,\gamma\delta}(\bm{q},\varepsilon,\omega) \\
    \nonumber
    &\quad= \gamma_0\delta_{\alpha,\gamma}\delta_{\beta,\delta}\\
    &\quad\quad+\gamma_0\sum_{\mu,\nu}\Pi^D_{\alpha\beta,\mu\nu}(\bm{q},\varepsilon,\omega)\Gamma^D_{\mu\nu,\gamma\delta}(\bm{q},\varepsilon,\omega),\\
    \label{eq:Bethe-Salpeter_Eq_Cooperon}
    \nonumber
    \displaybreak[0]
    &\Gamma^C_{\alpha\delta,\gamma\beta}(\bm{q},\varepsilon,\omega)\\
    \nonumber
    &\quad= \gamma_0^2\Pi^C_{\alpha\delta,\gamma\beta}(\bm{q},\varepsilon,\omega) \\
    &\quad\quad+ \gamma_0\sum_{\mu,\nu}\Pi^C_{\alpha\delta,\mu\nu}(\bm{q},\varepsilon,\omega)\Gamma^C_{\mu\nu,\gamma\beta}(\bm{q},\varepsilon,\omega),
\end{align}
hold. 
For the Cooperon, Eq.~(\ref{eq:Bethe-Salpeter_Eq_Cooperon}) is obtained by "twisting" the $\Gamma^C$ diagram in Fig.~\ref{fig:diagram}.
Here, $\Pi^D$ and $\Pi^C$ are
\begin{align}
    \label{eq:def_Pi^D}
    \nonumber
    &\Pi^D_{\alpha\beta,\gamma\delta}(\bm{q},\varepsilon,\omega)\\ 
    &= \int\frac{d^2\bm{k}''}{(2\pi)^2}\overline{\mathrm{G}}^R_{\alpha\gamma}\left(\bm{k}'',\varepsilon+\frac{\omega}{2}\right)\overline{\mathrm{G}}^A_{\delta\beta}\left(\bm{q}+\bm{k}'',\varepsilon-\frac{\omega}{2}\right),\\
    \label{eq:def_Pi^C}
    \nonumber
    &\Pi^C_{\alpha\beta,\gamma\delta}(\bm{q},\varepsilon,\omega)\\
    &= \int\frac{d^2\bm{k}''}{(2\pi)^2}\overline{\mathrm{G}}^R_{\alpha\gamma}\left(\bm{k}'',\varepsilon+\frac{\omega}{2}\right)\overline{\mathrm{G}}^A_{\beta\delta}\left(\bm{q}-\bm{k}'',\varepsilon-\frac{\omega}{2}\right),
\end{align}
where $\overline{\mathrm{G}}^{R/A}_{\alpha\beta}$ is the element of disorder-averaged Green's function in the spin basis,
\begin{widetext}
\begin{equation}
    \overline{\mathrm{G}}^{R/A}(\bm{q}, \varepsilon) = \frac12\left(
    \begin{array}{cc}
        g_+^{R/A}(\bm{q}, \varepsilon)+g_-^{R/A}(\bm{q}, \varepsilon) & -e^{-i\theta(\bm{q})}\left(g_+^{R/A}(\bm{q}, \varepsilon)-g_-^{R/A}(\bm{q}, \varepsilon)\right)\\[1mm]
        -e^{i\theta(\bm{q})}\left(g_+^{R/A}(\bm{q}, \varepsilon)-g_-^{R/A}(\bm{q}, \varepsilon)\right) & g_+^{R/A}(\bm{q}, \varepsilon)+g_-^{R/A}(\bm{q}, \varepsilon)
    \end{array}
    \right),
\end{equation}
and
\begin{equation}
    \gamma_0 = \frac{1}{\pi\rho\tau}.
\end{equation}
Here, $\rho$ is the density of states.
\begin{figure}[tbp]
  \includegraphics[width=0.5\linewidth]{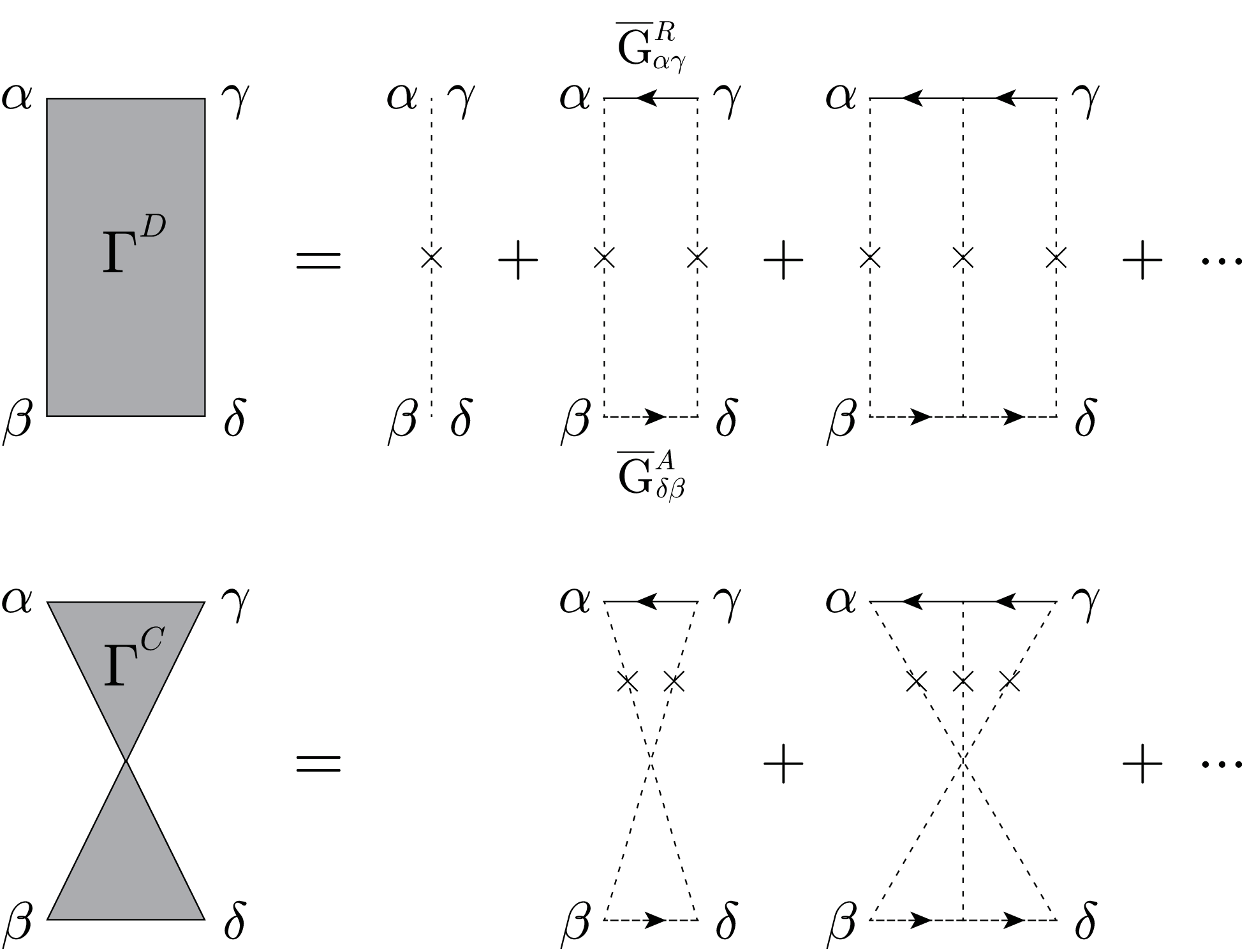}
  \caption{\label{fig:diagram}
  Diffuson and Cooperon diagrams. 
  The solid line arrows and dashed line arrows represent the disorder-averaged retarded Green's function and advanced Green's function, respectively. 
  The cross marks and dotted lines represent interactions with impurities. 
  The subscripts correspond to Eqs.~(\ref{eq:Bethe-Salpeter_Eq_Diffuson}) and (\ref{eq:Bethe-Salpeter_Eq_Cooperon}).
  }
\end{figure}

When considering $\bm{k}=-\bm{k}_0$ in Eq.~(\ref{eq:n_initial_state_k0_+}), it is apparent that by using Eqs.~(\ref{eq:transform_Gamma^D_spin_band}) and (\ref{eq:transform_Gamma^C_spin_band}), it is sufficient to calculate $\Gamma^D(\bm{0},\varepsilon,\omega)$ and $\Gamma^C(\bm{0},\varepsilon,\omega)$, or equivalently $\Pi^D(\bm{0},\varepsilon,\omega)$ and $\Pi^C(\bm{0},\varepsilon,\omega)$. 
Performing the integration in Eqs.~(\ref{eq:def_Pi^D}) and (\ref{eq:def_Pi^C}) taking into account factors such as $e^{i\theta(-\bm{k})}=-e^{i\theta(\bm{k})}$ and $e^{2i\theta(k_x,k_y)}=-e^{2i\theta(k_y,-k_x)}$, several terms cancel out, and we obtain
\begin{align}
    \displaybreak[0]
    \label{eq:Pi^D_matrix}
    \Pi^D(\bm{0},\varepsilon,\omega) &= \left(
    \begin{array}{cccc}
        \Pi_1 & 0 & 0 & \Pi_2 \\
        0 & \Pi_1 & 0 & 0 \\
        0 & 0 & \Pi_1 & 0 \\
        \Pi_2 & 0 & 0 & \Pi_1
    \end{array}
    \right),\\
    \displaybreak[0]
    \label{eq:Pi^C_matrix}
    \Pi^C(\bm{0},\varepsilon,\omega) &= \left(
    \begin{array}{cccc}
        \Pi_1 & 0 & 0 & 0 \\
        0 & \Pi_1 & -\Pi_2 & 0 \\
        0 & -\Pi_2 & \Pi_1 & 0 \\
        0 & 0 & 0 & \Pi_1
    \end{array}
    \right),
\end{align}
where
\begin{align}
    \displaybreak[0]
    \label{eq:def_Pi1}
    \Pi_1(\varepsilon,\omega) &= \frac14\int\frac{d^2\bm{k}''}{(2\pi)^2}\left[g_+^R\left(\bm{k}'',\varepsilon+\frac{\omega}{2}\right)+g_-^R\left(\bm{k}'',\varepsilon+\frac{\omega}{2}\right)\right]\left[g_+^A\left(\bm{k}'',\varepsilon-\frac{\omega}{2}\right)+g_-^A\left(\bm{k}'',\varepsilon-\frac{\omega}{2}\right)\right],\\
    \displaybreak[0]
    \label{eq:def_Pi2}
    \Pi_2(\varepsilon,\omega) &= \frac14\int\frac{d^2\bm{k}''}{(2\pi)^2}\left[g_+^R\left(\bm{k}'',\varepsilon+\frac{\omega}{2}\right)-g_-^R\left(\bm{k}'',\varepsilon+\frac{\omega}{2}\right)\right]\left[g_+^A\left(\bm{k}'',\varepsilon-\frac{\omega}{2}\right)-g_-^A\left(\bm{k}'',\varepsilon-\frac{\omega}{2}\right)\right].
\end{align}
Substituting Eqs.~(\ref{eq:Pi^D_matrix})and (\ref{eq:Pi^C_matrix}) into the Bethe-Salpeter equations~(\ref{eq:Bethe-Salpeter_Eq_Diffuson}) and (\ref{eq:Bethe-Salpeter_Eq_Cooperon}), we obtain
\begin{align}
    \displaybreak[0]
    \label{eq:Gamma^D_matrix}
    \Gamma^D(\bm{0},\varepsilon,\omega) &= \left(
    \begin{array}{cccc}
        \frac{\Gamma_1+\Gamma_3}{2} & 0 & 0 & \frac{\Gamma_1-\Gamma_3}{2} \\
        0 & \Gamma_2 & 0 & 0 \\
        0 & 0 & \Gamma_2 & 0 \\
        \frac{\Gamma_1-\Gamma_3}{2} & 0 & 0 & \frac{\Gamma_1+\Gamma_3}{2}
    \end{array}
    \right),\\
    \displaybreak[0]
    \label{eq:Gamma^C_matrix}
    \Gamma^C(\bm{0},\varepsilon,\omega) &= \left(
    \begin{array}{cccc}
        \Gamma_2-\gamma_0 & 0 & 0 & -\frac{\Gamma_1-\Gamma_3}{2} \\
        0 & \frac{\Gamma_1+\Gamma_3}{2}-\gamma_0 & 0 & 0 \\
        0 & 0 & \frac{\Gamma_1+\Gamma_3}{2}-\gamma_0 & 0 \\
        -\frac{\Gamma_1-\Gamma_3}{2} & 0 & 0 & \Gamma_2-\gamma_0
    \end{array}
    \right),
\end{align}
where
\begin{align}
    \displaybreak[0]
    \label{eq:def_Gamma1}
    \Gamma_1(\varepsilon,\omega) &= \frac{\gamma_0}{1-\gamma_0\left[\Pi_1(\varepsilon,\omega)+\Pi_2(\varepsilon,\omega)\right]},\\
    \displaybreak[0]
    \label{eq:def_Gamma2}
    \Gamma_2(\varepsilon,\omega) &= \frac{\gamma_0}{1-\gamma_0\Pi_1(\varepsilon,\omega)},\\
    \label{eq:def_Gamma3}
    \displaybreak[0]
    \Gamma_3(\varepsilon,\omega) &= \frac{\gamma_0}{1-\gamma_0\left[\Pi_1(\varepsilon,\omega)-\Pi_2(\varepsilon,\omega)\right]}.
\end{align}
By transforming $\Gamma^D$ and $\Gamma^C$ into the $\pm$~basis using Eqs.~(\ref{eq:transform_Gamma^D_spin_band}) and (\ref{eq:transform_Gamma^C_spin_band}) respectively, and substituting them into Eqs.~(\ref{eq:Phi^D}) and (\ref{eq:Phi^C}), and rearranging Eq.~(\ref{eq:n_initial_state_k0_+}), we obtain the disorder-averaged momentum distribution at $\bm{k}=-\bm{k}_0$,
\begin{align}
    \nonumber
    \label{eq:n_CBS_direction}
    n_-(-\bm{k}_0, t) &= \int_{-\infty}^{\infty}\frac{d\varepsilon}{2\pi}\int_{-\infty}^{\infty}\frac{d\omega}{2\pi}e^{-i\omega t}\left[\frac{\Gamma_1(\varepsilon,\omega)}{2} + \Gamma_2(\varepsilon,\omega) + \frac{\Gamma_3(\varepsilon,\omega)}{2} - \gamma_0\right]\\
    &\quad\times g_+^R\left(\bm{k}_0,\varepsilon+\frac{\omega}{2}\right)g_+^A\left(\bm{k}_0,\varepsilon-\frac{\omega}{2}\right)g_-^R\left(\bm{k}_0,\varepsilon+\frac{\omega}{2}\right)g_-^A\left(\bm{k}_0,\varepsilon-\frac{\omega}{2}\right),\\
    n_+(-\bm{k}_0, t) &= 0.
\end{align}
Since $|\bm{k}_0,+\rangle$ and $|-\bm{k}_0,+\rangle$ are time-reversed states, $n_+(-\bm{k}_0)=0$, which is consistent with Eq.~(\ref{eq:no-scattering_to_time_reversed_state}).

\begin{figure}[tbp]
  \includegraphics[width=0.75\linewidth]{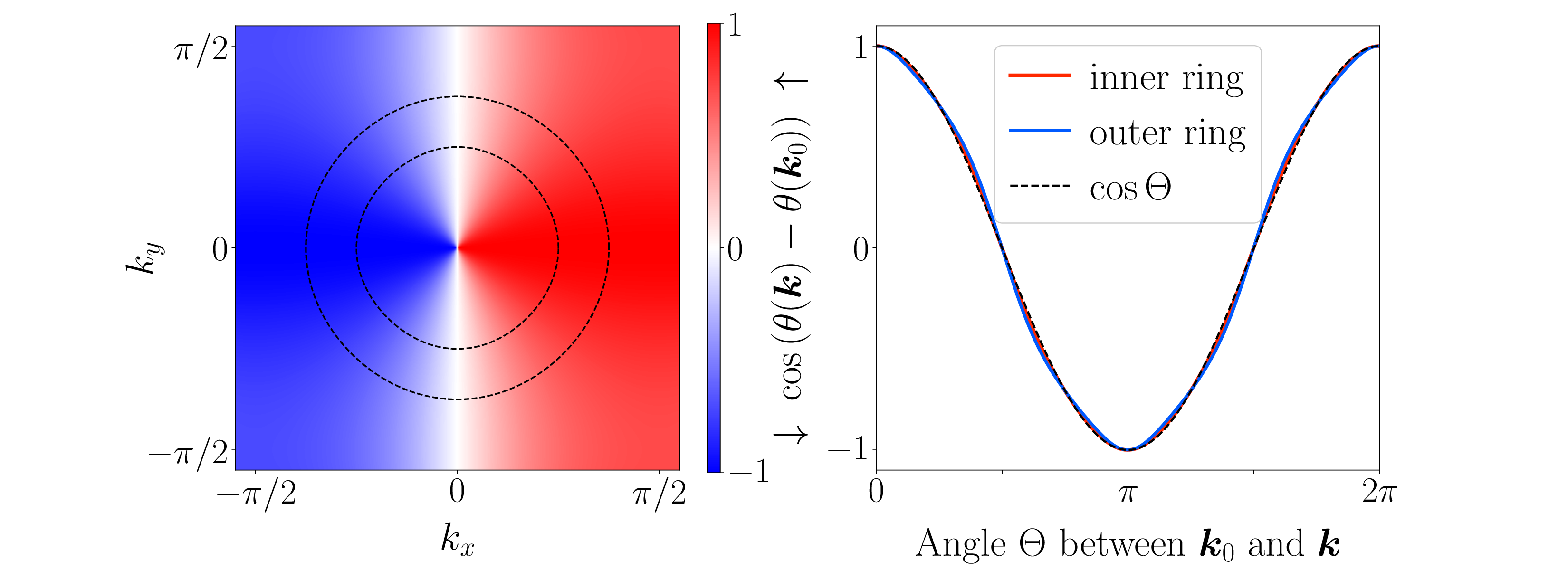}
  \caption{\label{fig:angle_vs_cos}
  (Color online) (left) The value of $\cos\big(\theta(\bm{k})-\theta(\bm{k}_0)\big)$ for $\bm{k}_0=(\pi/4,0)$. 
  The dashed lines represent energy contours.
  (right) The value of $\cos\big(\theta(\bm{k})-\theta(\bm{k}_0)\big)$ on the inner and outer rings shown as dashed lines in the left figure.
  We compare with $\cos\Theta$ where $\Theta$ is the angle between $\bm{k}$ and $\bm{k}_0$.
  }
\end{figure}
If $\bm{k}$ is not close to $-\bm{k}_0$, the contribution of the Cooperon is negligible. In this case, the disorder-averaged momentum distribution at $\bm{k}$ is 
\begin{align}
    \nonumber
    n_{\pm}(\bm{k}, t) &= \int_{-\infty}^{\infty}\frac{d\varepsilon}{2\pi}\int_{-\infty}^{\infty}\frac{d\omega}{2\pi}e^{-i\omega t}\left[\Gamma_1(\varepsilon,\omega) \pm \Gamma_2(\varepsilon,\omega)\cos\big(\theta(\bm{k})-\theta(\bm{k}_0)\big)\right]\\
    \label{eq:n_other_direction}
    &\quad\times g_+^R\left(\bm{k}_0,\varepsilon+\frac{\omega}{2}\right)g_+^A\left(\bm{k}_0,\varepsilon-\frac{\omega}{2}\right)g_{\pm}^R\left(\bm{k},\varepsilon+\frac{\omega}{2}\right)g_{\pm}^A\left(\bm{k},\varepsilon-\frac{\omega}{2}\right).
\end{align}
\end{widetext}
Here, for the moment we ignore the terms associated with $\mathit{\Phi}^0_{\pm}$ since they are zero except for $\bm{k}=\bm{k}_0$.
The first term in Eq.~(\ref{eq:n_other_direction}) is an isotropic contribution, while the second term is an anisotropic contribution proportional to $\cos(\theta(\bm{k})-\theta(\bm{k}_0))$. 
The left panel of Fig.~\ref{fig:angle_vs_cos} shows how $\cos\big(\theta(\bm{k})-\theta(\bm{k}_0)\big)$ varies with $\bm{k}$ when $\bm{k}_0=(\pi/4,0)$. 
The dashed curves are energy contours with energy $E_+(\bm{k}_0)$ for the Ando model without disorder ($W=0$) for SOC strength $\varphi=\pi/16$. 
In the right panel of Fig.~\ref{fig:angle_vs_cos}, we show the variation of the factor $\cos\big(\theta(\bm{k})-\theta(\bm{k}_0)\big)$ along these energy contours. 
We see that, for the parameters considered in this paper, this factor is well approximated by $\cos\Theta$ where $\Theta$ is the angle between $\bm{k}_0$ and $\bm{k}$.

We define $\Delta$ so that $\Delta = E_+(\bm{k}_0) - E_-(\bm{k}_0)$, and approximate $\Gamma_1$, $\Gamma_2$, and $\Gamma_3$ by evaluating $\Pi_1$ and $\Pi_2$ in Eqs.~(\ref{eq:def_Gamma1})-(\ref{eq:def_Gamma3}) as
\begin{align}
    \displaybreak[0]
    \label{eq:approx_Gamma_1}
    \Gamma_1(\varepsilon,\omega) &\simeq \gamma_0\frac{\omega+\frac{i}{\tau}}{\omega},\\
    \displaybreak[0]
    \label{eq:approx_Gamma_2}
    \Gamma_2(\varepsilon,\omega) &\simeq \gamma_0\frac{\left(\omega+\frac{i}{\tau}\right)\left(\omega+\Delta+\frac{i}{\tau}\right)\left(\omega-\Delta+\frac{i}{\tau}\right)}{(\omega+i\omega_1)(\omega+i\omega_2)(\omega+i\omega_3)},\\
    \label{eq:approx_Gamma_3}
    \Gamma_3(\varepsilon,\omega) &\simeq \gamma_0\frac{\left(\omega+\Delta+\frac{i}{\tau}\right)\left(\omega-\Delta+\frac{i}{\tau}\right)}{(\omega+i\omega_4)(\omega+i\omega_5)},
\end{align}
where $\omega_1,\ \omega_2,\ \omega_3$ are the roots of the following cubic equation,
\begin{equation}\label{eq:cubic_eq}
    \omega^3 - \frac{2}{\tau}\omega^2 + \left(\frac{1}{\tau^2}+\Delta^2\right)\omega -\frac{\Delta^2}{2\tau} = 0,
\end{equation}
and $\omega_4,\ \omega_5$ are the roots of the quadratic equation,
\begin{equation}\label{eq:quadratic_eq}
    \omega^2 - \frac{1}{\tau}\omega +\Delta^2 = 0.
\end{equation}
These equations were also obtained in Ref.~\cite{Szolnoki_2017_srep}.
Note that both $\tau\omega_l$ and $\omega_l/\Delta$ ($l=1,2,\cdots,5$) can be expressed as functions of a single variable, $\tau\Delta$.
Detailed calculations are given in Appendix~\ref{app:approx_Gamma}, but we emphasise that the method we use is not perturbative in the strength of the SOC, i.e., Eqs.~(\ref{eq:approx_Gamma_1})-(\ref{eq:approx_Gamma_3}) are good approximation even if $\Delta$ is larger than $1/\tau$, provided $\Delta$ is sufficiently smaller than the band width. 
The term $\Gamma_1$ in Eq.~(\ref{eq:approx_Gamma_1}) has a singularity at $\omega=0$. Therefore, after a sufficiently long time, only the contribution from $\Gamma_1$ remains.
Equation~(\ref{eq:cubic_eq}) has one real root, which we denote by $\omega_1$, and two complex conjugate roots, which we denote by $\omega_2$ and $\omega_3$. All the roots have positive real parts. 
The two roots $\omega_4$ and $\omega_5$ of Eq.~(\ref{eq:quadratic_eq}) are real when $2\tau\Delta<1$, and complex conjugates when $2\tau\Delta>1$. The real parts of these roots are always positive. 
We summarise the properties of the roots in Table~\ref{tab:roots_of_equation} for the limits of weak and strong SOC. 
In the case of weak SOC, energy scales of $\tau\Delta^2$ and $1/\tau$ appear, while in the case of strong SOC, energy scales of $\Delta$ and $1/\tau$ appear. 
These energies correspond to the time scales tabulated in Table~\ref{tab:time_scale}.

\begin{table}[t]
    \caption{
    The roots of Eqs.~(\ref{eq:cubic_eq}) and (\ref{eq:quadratic_eq}) in the weak and strong limits of the SOC. 
    Note that we are setting $\hbar=1$ in this section.
    \label{tab:roots_of_equation}}
    \begin{ruledtabular}
    \renewcommand{\arraystretch}{1.2}
    \begin{tabular}{ccc}
    & weak SOC & strong SOC \\
     &$\tau\Delta\ll1$&$\tau\Delta\gg1$\\
    \hline
    \rule{0pt}{1.7\normalbaselineskip}$\omega_1$& $\displaystyle \frac{\tau\Delta^2}{2}$ & $\displaystyle \frac{1}{2\tau}$ \\[9pt]
    $\omega_2$, $\omega_3$& $\displaystyle \frac{1}{\tau}\pm i\frac{\Delta}{\sqrt{2}}$ & $\displaystyle \frac{3}{4\tau}\pm i\Delta$ \\[9pt]
    $\omega_4$, $\omega_5$& $\displaystyle \tau\Delta^2,\ \frac{1}{\tau}\big(1-(\tau\Delta)^2\big)$ & $\displaystyle \frac{1}{2\tau}\pm i\Delta$ \\[4pt]
    \end{tabular}
    \end{ruledtabular}
\end{table}

Substituting Eqs.~(\ref{eq:approx_Gamma_1})-(\ref{eq:approx_Gamma_3}) into Eq.~(\ref{eq:n_CBS_direction}), and using the residue theorem, we obtain the disorder-averaged momentum distribution at $\bm{k}=-\bm{k}_0$,
\begin{widetext}
\begin{equation}\label{eq:analytic_n_CBS_weak_SOC}
    n_-(-\bm{k}_0,t) = f\left(t,0, \frac{1}{\tau}, \Delta\right)+ 2f\left(t,\omega_1, \frac{\omega_2+\omega_3}{2}, \frac{\omega_2-\omega_3}{2i}\right) + g\left(t,\frac{1}{\tau}, \omega_4, \omega_5\right) - 2f\left(t,\frac{1}{\tau}, \frac{1}{\tau}, \Delta\right),
\end{equation}
for $2\tau\Delta<1$ and
\begin{equation}\label{eq:analytic_n_CBS_strong_SOC}
    n_-(-\bm{k}_0,t) = f\left(t,0, \frac{1}{\tau}, \Delta\right)+ 2f\left(t,\omega_1, \frac{\omega_2+\omega_3}{2}, \frac{\omega_2-\omega_3}{2i}\right)+ f\left(t,\frac{1}{\tau}, \frac{\omega_4+\omega_5}{2}, \frac{\omega_4-\omega_5}{2i}\right) - 2f\left(t,\frac{1}{\tau}, \frac{1}{\tau}, \Delta\right),
\end{equation}
for $2\tau\Delta>1$, where 
\begin{align}
    \displaybreak[0]
    f(t,x,y,z) &= \frac{\gamma_0}{(x-y)^2+z^2}\left[e^{-xt} - e^{-yt}\left(\cos(zt)-\frac{x-y}{z}\sin(zt)\right)\right],\\
    g(t,x,y,z) &= -\frac{\gamma_0}{(x-y)(y-z)(z-x)}\Big[(y-z)e^{-xt} + (z-x)e^{-yt} + (x-y)e^{-zt}\Big].
\end{align}
The first terms of Eqs.~(\ref{eq:analytic_n_CBS_weak_SOC}) and (\ref{eq:analytic_n_CBS_strong_SOC}) remain at $t\to\infty$. On the other hand, substituting Eqs.~(\ref{eq:approx_Gamma_1})-(\ref{eq:approx_Gamma_3}) into Eq.~(\ref{eq:n_other_direction}), we obtain
\begin{equation}\label{eq:analytic_n_other_direction}
    n_{\pm}(\bm{k},t) = f\left(t,0, \frac{1}{\tau}, \tilde{\Delta}_{\pm}(\bm{k})\right)\pm h\left(t,\omega_1, \frac{\omega_2+\omega_3}{2}, \frac{\omega_2-\omega_3}{2i},\frac{1}{\tau},\tilde{\Delta}_{\pm}(\bm{k}),\frac{1}{\tau},\Delta\right)\cos\big(\theta(\bm{k})-\theta(\bm{k}_0)\big),
\end{equation}
where
\begin{equation}
    \tilde{\Delta}_{\pm}(\bm{k}) = E_+(\bm{k}_0) - E_{\pm}(\bm{k}),
\end{equation}
and 
\begin{align}\label{eq:function_h}
    \nonumber
    h(&t,s,u,v,w,x,y,z) \\
    \nonumber
    \displaybreak[0]
    &= \frac{\gamma_0e^{-st}\left\{(s-y)^2+z^2\right\}}{\left\{(s-u)^2+v^2\right\}\left\{(s-w)^2+x^2\right\}}\\
    \displaybreak[0]
    \nonumber
    &\quad -\gamma_0e^{-ut}\left[\frac{\left\{i(y-u)+(z+v)\right\}\left\{i(y-u)-(z-v)\right\}e^{-ivt}}{2v\left\{i(s-u)+v\right\}\left\{i(w-u)+(x+v)\right\}\left\{i(w-u)-(x-v)\right\}}+h.c.\right]\\
    \displaybreak[0]
    &\quad -\gamma_0e^{-wt}\left[\frac{\left\{i(y-w)+(z+x)\right\}\left\{i(y-w)-(z-x)\right\}e^{-ixt}}{2x\left\{i(s-w)+x\right\}\left\{i(u-w)+(v+x)\right\}\left\{i(u-w)-(v-x)\right\}}+h.c.\right].
\end{align}
Also, when $\bm{k}=\bm{k}_0$, there is an additional contribution
\begin{equation}
    \int_{-\infty}^{\infty}\frac{d\varepsilon}{2\pi}\int_{-\infty}^{\infty}\frac{d\omega}{2\pi}e^{-i\omega t}\mathit{\Phi}^0_{\pm}(\bm{k},\varepsilon,\omega) = e^{-\frac{t}{\tau}}\delta(\bm{k}-\bm{k}_0)\delta_{\pm,+}.
\end{equation}
The discussion up to this point can be carried out in the same manner when the initial state is $|\bm{k}_0,-\rangle$.

The diagrammatic calculation presented above is valid for weak weak disorder. More precisely, the condition is
that $k_\mathrm{F} l_e \gg 1$, where $k_\mathrm{F}$ is the Fermi wave vector and $l_e$ is the mean free path.
In addition, the time scales under consideration should be not too long.
The relevant time scales here are the Thouless time $\tau_D = L^2 / D$, where $L$ is the system size and $D = v^2 \tau / 2$ is the diffusion constant (with $v$ the group velocity), and the localisation time  $\tau_{\rm loc} = \xi^2 / D$ 
where $\xi$ is the localisation length.
For times longer than the shorter of these two time scales, a coherent forward scattering peak~\cite{Karpiuk_2012} is expected.
The Diffuson and the Cooperon are not sufficient to capture this phenomenon.
\end{widetext}

\section{\label{sec:result}Results}

We performed simulations of the time evolution of a wave packet using the 2D Ando model~(\ref{eq:Hamiltonian_Ando_real_spin_basis}) defined on an $L\times L$ lattice with $L=512$ and periodic boundary conditions.
We used the Chebyshev expansion method~\cite{Tal-Ezer_1984} to calculate the time evolution of the state vector.
The initial state was a plane wave~(\ref{eq:initial_state}) with a wave vector of $\bm{k}_0 = (\pi/4,0)$. 
We considered three different values for the Ando model parameter $\varphi$:~$\pi/1024$, $\pi/128$, and $\pi/16$. 
In each case, we fixed the disorder strength at $W = 1$ and estimated the average of the momentum distribution by sampling $2^{13} = 8192$ disorder realisations.
We used the coherent potential approximation~\cite{Soven_1967} to estimate the density of states, scattering time, and mean free
path for these parameters.
The values are given in Table~\ref{tab:Ando_model_parameters} together with the spin splitting $\Delta$ and the value of the dimensionless
ratio $\tau \Delta / \hbar$.
\begin{table}[b]
    \caption{
    The scattering time, spin splitting, and density of states per unit area for $\varphi=\pi/1024$, $\varphi=\pi/128$, and $\varphi=\pi/16$. 
    These values are used to calculate the solid lines in Fig.~\ref{fig:time_dependence_CBS}(d)-(f).
    The lattice constant is the unit of length, $t_{\mathrm{hop}}$ in Eq.~(\ref{eq:Hamiltonian_Ando_real_spin_basis}) is the unit of energy, and $\hbar/t_{\mathrm{hop}}$ is the unit of time.
    The mean free path $l_e=v\tau$, where $v$ is the group velocity, is not used in the calculation.
    We note that $k_0l_e\gg1$ for the parameters considered in our simulations.
    \label{tab:Ando_model_parameters}}
    \begin{ruledtabular}
    \renewcommand{\arraystretch}{1.2}
    \begin{tabular}{cccc}
    $\varphi$ &$\pi/1024$&$\pi/128$&$\pi/16$\\
    \hline
    SOC strength $\tau\Delta/\hbar$ & $0.194$ & $1.55$ & $11.8$\\
    spin splitting $\Delta$ & $8.68\times10^{-3}$ & $0.0694$ & $0.552$ \\
    scattering time $\tau$ & $22.34$ & $22.40$ & $21.31$\\
    mean free path $l_e$ & $31.7$ & $32.4$ & $35.4$\\
    density of states $\rho$ & $0.1713$ & $0.1721$ & $0.1806$\\
    \end{tabular}
    \end{ruledtabular}
\end{table}

We focus first on the time dependence of the disorder-averaged momentum distribution $n(\bm{k},t)$ at $\bm{k}=-\bm{k}_0$, which is
parallel to the backscattering direction.
In Fig.~\ref{fig:time_dependence_CBS}(a)-(c), we plot $n(-\bm{k}_0, t)$ obtained from our simulations.
For comparison we also plot $n(\bm{k}, t)$ at $\bm{k}=(0,\pi/4)$, which is orthogonal to the backscattering direction.
For weak SOC (Fig.~\ref{fig:time_dependence_CBS}(a)), $n(-\bm{k}_0,t)$ increases from zero and peaks at approximately twice the diffusive background 
on a time scale of the order of several scattering times, i.e., for short times this behaviour is similar to CBS observed in the absence of SOC.
After this, however, there is a crossover to decreasing behaviour with a limiting value of $n(-\bm{k}_0,t)$ equal to
half the diffusive background at long times.
For intermediate SOC strength (Fig.~\ref{fig:time_dependence_CBS}(b)), we see that $n(-\bm{k}_0,t)$ still exhibits
a maximum at short times and at longer times tends to a limiting value that is a small fraction of the diffusive background. 
For strong SOC (Fig.~\ref{fig:time_dependence_CBS}(c)), we see that $n(-\bm{k}_0,t)$ is strongly reduced compared to the diffusive background, exhibits decaying oscillations, and tends to a constant value at long times.

In Fig.~\ref{fig:time_dependence_CBS}(d)-(f), we overlay the disorder-averaged momentum distribution at $\bm{k}=-\bm{k}_0$ obtained from simulations with calculations of Eq.~(\ref{eq:analytic_n_CBS_weak_SOC}) or (\ref{eq:analytic_n_CBS_strong_SOC}) with values of
the scattering time $\tau$ and the spin splitting $\Delta$ given in Table~\ref{tab:Ando_model_parameters}.
The analytical equations and the simulation results exhibit a clear and consistent agreement, for weak, intermediate, and strong SOC. This demonstrates that  considering only the Diffuson and the Cooperon provides a good approximation for the disorder-averaged momentum distribution.

\begin{figure}[tbp]
  \includegraphics[width=0.8\linewidth]{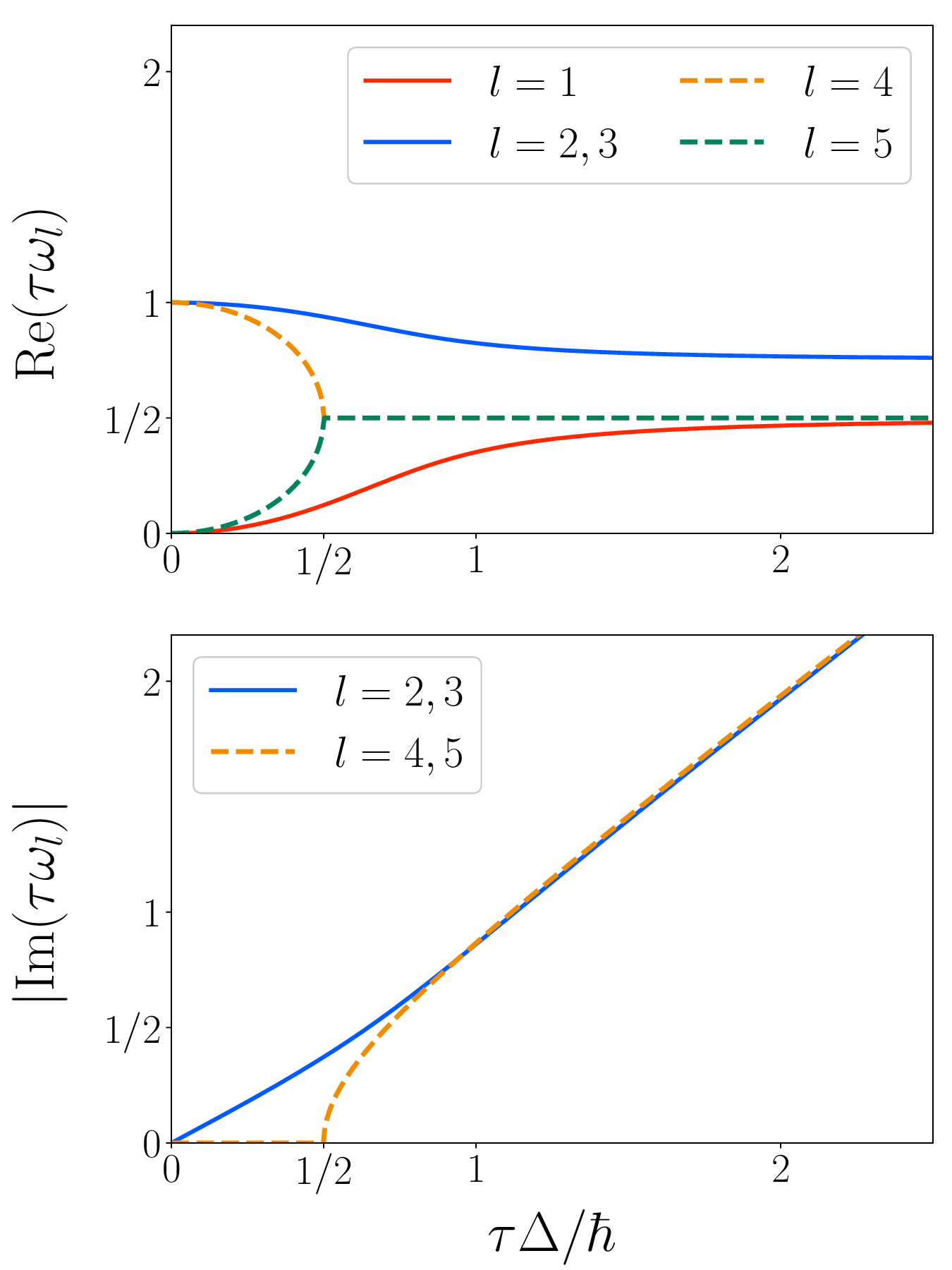}
  \caption{\label{fig:roots_of_equations}
  (Color online) SOC strength $\Delta$ dependence of the roots $\omega_l$ ($l=1,2\cdots,5$) of Eqs.~(\ref{eq:cubic_eq}) and (\ref{eq:quadratic_eq}).
  The real part of $\omega_l$ is shown at the top and the absolute value of the imaginary part at the bottom.
  The solid lines represent the roots of Eq.~(\ref{eq:cubic_eq}) ($\mathrm{Re}(\omega_1)<\mathrm{Re}(\omega_2)=\mathrm{Re}(\omega_3)$), and the dashed lines represent the roots of Eq.~(\ref{eq:quadratic_eq}).
  For convenience, we define $\mathrm{Re}(\omega_4) \geq \mathrm{Re}(\omega_5)$.
  We make the abscissa and ordinate dimensionless by multiplying by the scattering time $\tau$. 
  Note that $\tau\omega_l$ is a function of a single variable, $\tau\Delta/\hbar$.}
\end{figure}

The time scales governing the behaviour of $n(-\bm{k}_0, t)$ are determined by the roots $\omega_l$ ($l=1,2,\cdots,5$) 
of Eqs.~(\ref{eq:cubic_eq}) and (\ref{eq:quadratic_eq}). 
In Fig.~\ref{fig:roots_of_equations}, we show the SOC strength $\Delta$ dependence of the roots.
From Eqs.~(\ref{eq:analytic_n_CBS_weak_SOC}) and (\ref{eq:analytic_n_CBS_strong_SOC}), 
it is evident that the reciprocals of the real parts of the roots determine the overall time dependence of
$n(-\bm{k}_0, t)$. 
If a root is a complex number, it signifies that the associated terms exhibit damped oscillations
with the imaginary part of the root giving the associated frequency of oscillation.
The term that gives the dominant contribution varies with the parameters of the system and with time.
To understand qualitatively the behaviour of $n(-\bm{k}_0, t)$, we focus on two extremes: the limit where the SOC is weak compared to disorder and the limit where it is strong.

\begin{figure}[tbp]
  \includegraphics[width=0.93\linewidth]{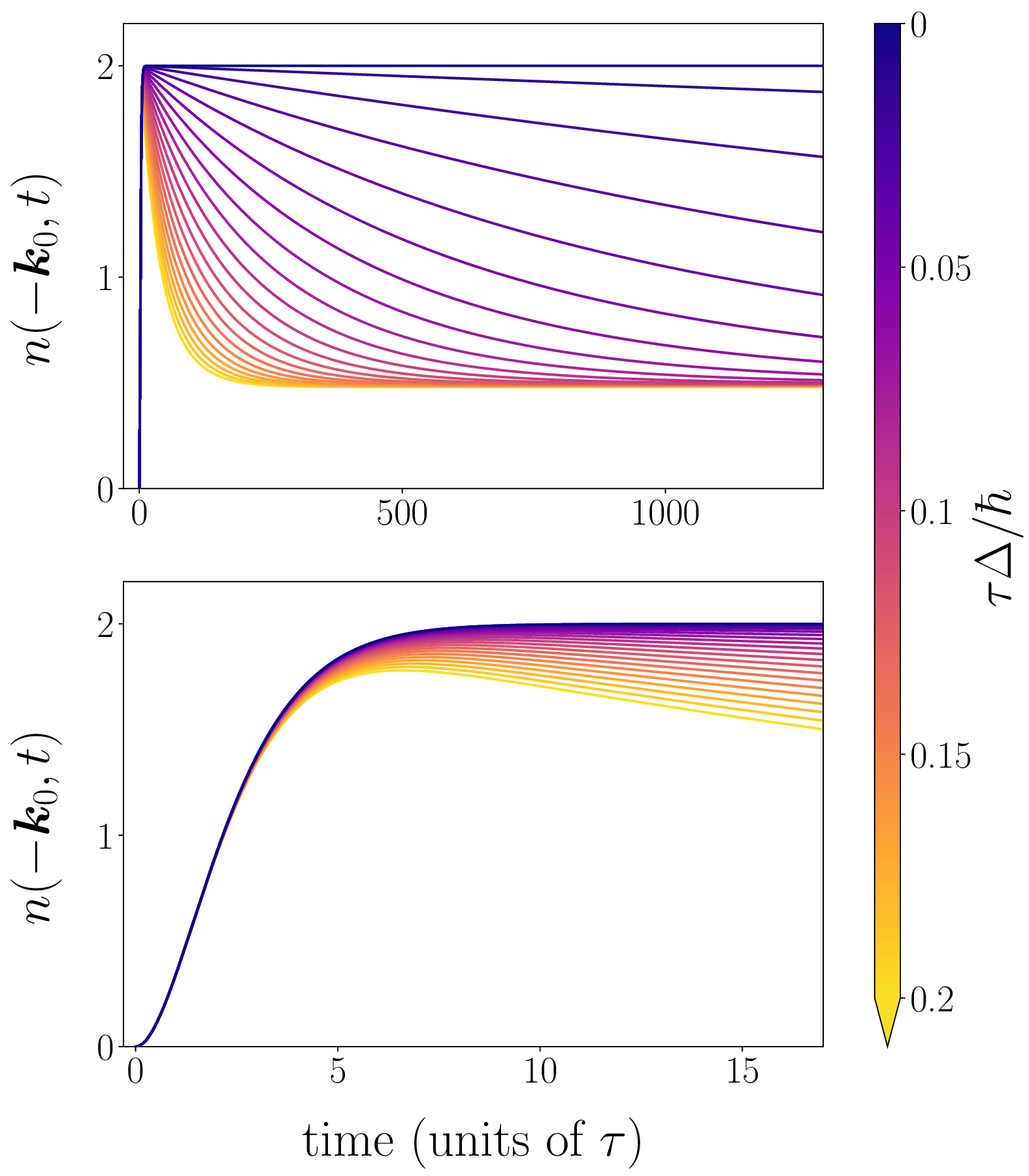}
  \caption{\label{fig:CBS_SOC-dependence_weak}
  (Color online) 
  The time evolution of the disorder-averaged momentum distribution at $\bm{k}=-\bm{k}_0$ for various values of $\tau\Delta/\hbar$ (from $0$ to $0.2$ with interval $0.01$) in the weak SOC regime obtained from 
  the diagrammatic calculation, i.e., Eq.~(\ref{eq:analytic_n_CBS_weak_SOC}).
  The short-time behaviour is shown in the bottom figure.
  The disorder-averaged momentum distribution is expressed in units of $2\tau/\pi\hbar\rho$, 
  the height of the diffusive background in the absence of SOC.
  This value differs by a factor of two corresponding to the spin degrees of freedom compared to that of Ref.~\cite{Cherroret_2012_PRA}.
  }
\end{figure}

For weak SOC, the enhancement of backscattering by a factor two compared to the diffusive background appears at a time scale of several $\tau$, 
while the subsequent reduction to one half of the diffusive background, that is shown in Fig.~\ref{fig:CBS_SOC-dependence_weak}, 
occurs on a much longer time scale of 
\begin{equation}
    \frac{\hbar^2}{\tau\Delta^2} = \left( \frac{\tau_p}{\tau} \right)^2 \tau ,
\end{equation}
where 
\begin{equation}
    \tau_p = \frac{\hbar}{\Delta}
\end{equation}
is the spin precession time.

For strong SOC, two diffusive rings appear because of the spin splitting due to the SOC.
The width of the rings, which is of order $\hbar/\tau$, is small compared with their splitting, which is of order 
$\delta k \approx \Delta /(\hbar v)$ where $v$ is the average of the group velocities at $\bm{k}_0$.
The oscillations that are shown in Fig.~\ref{fig:time_dependence_CBS}(f) arise from the Diffuson and Cooperon contributions to
$n_-(-\bm{k}_0, t)$.
At $\bm{k}=-\bm{k}_0$, $n_+(-\bm{k}_0,t)=0$ holds exactly for all $t$.
The period of oscillation coincides with the spin precession time $\tau_p$.

\begin{figure}[tbp]
  \includegraphics[width=0.8\linewidth]{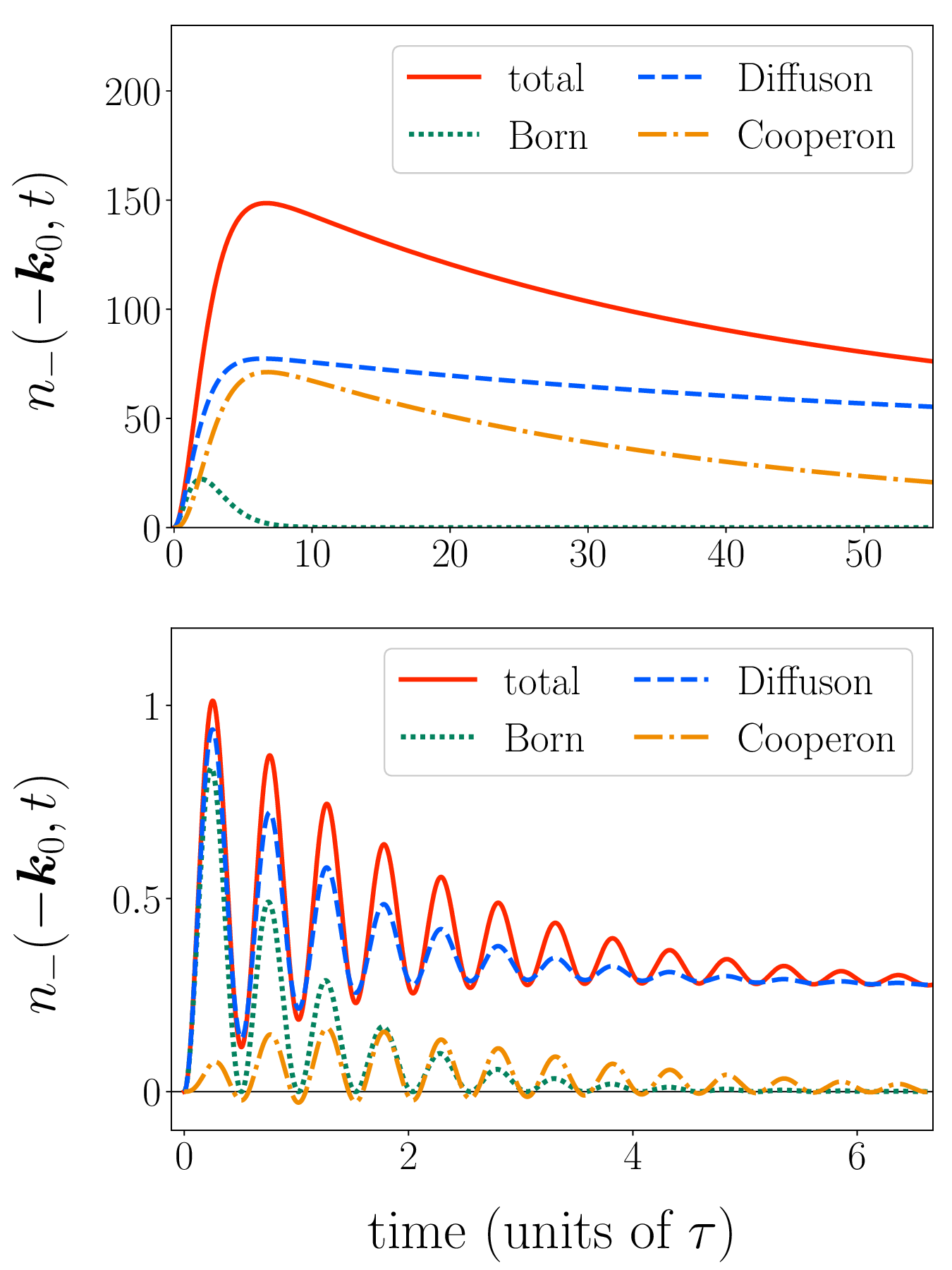}
  \caption{\label{fig:Born_Diffuson_Cooperon}
  (Color online) The disorder-averaged momentum distribution at $\bm{k}=-\bm{k}_0$ decomposed into the contributions from single scattering (Born approximation), Diffuson, and Cooperon. 
  Note we define the Diffuson to include the lowest-order diagram.
  The solid lines labeled "total" correspond to the solid lines in Fig.~\ref{fig:time_dependence_CBS}(d) and (f).
  (top)~$\varphi=\pi/1024$ (bottom)~$\varphi=\pi/16$.
  }
\end{figure}

While the contributions of the Diffuson and Cooperon at $\bm{k}=-\bm{k}_0$ cannot be separated in the simulation, 
they can be separated analytically.
The contributions of the Diffuson and Cooperon to $n_+(-\bm{k}_0,t)$ are equal in magnitude but opposite in sign, and cancel exactly.
In Fig.~\ref{fig:Born_Diffuson_Cooperon}, we investigate the contributions to $n_-(-\bm{k}_0,t)$ 
of the Diffuson and Cooperon for both weak and strong SOC.
In both cases, the short-time behaviour can be explained by the expression derived from the single scattering diagram 
(Born approximation). 
However, beyond the scattering time $\tau$, the contribution of multiple scattering becomes important. 
Oscillations of $n_-(-\bm{k}_0,t)$ gradually decay, but the oscillations originating from the Cooperon persist longer 
than those originating from the Diffuson.

We now shift our attention to the time dependence of the disorder-averaged momentum distribution at other momenta $\bm{k}\neq-\bm{k}_0$. 
The initial state is spin polarised and there is consequently a resulting anisotropy in
the disorder-averaged momentum distribution.
This relaxes to an isotropic distribution only at sufficiently long times as a result of spin relaxation.
By investigating this we determine the spin relaxation time for 
both weak and strong SOC.

\begin{figure}[tbp]
  \includegraphics[width=\linewidth]{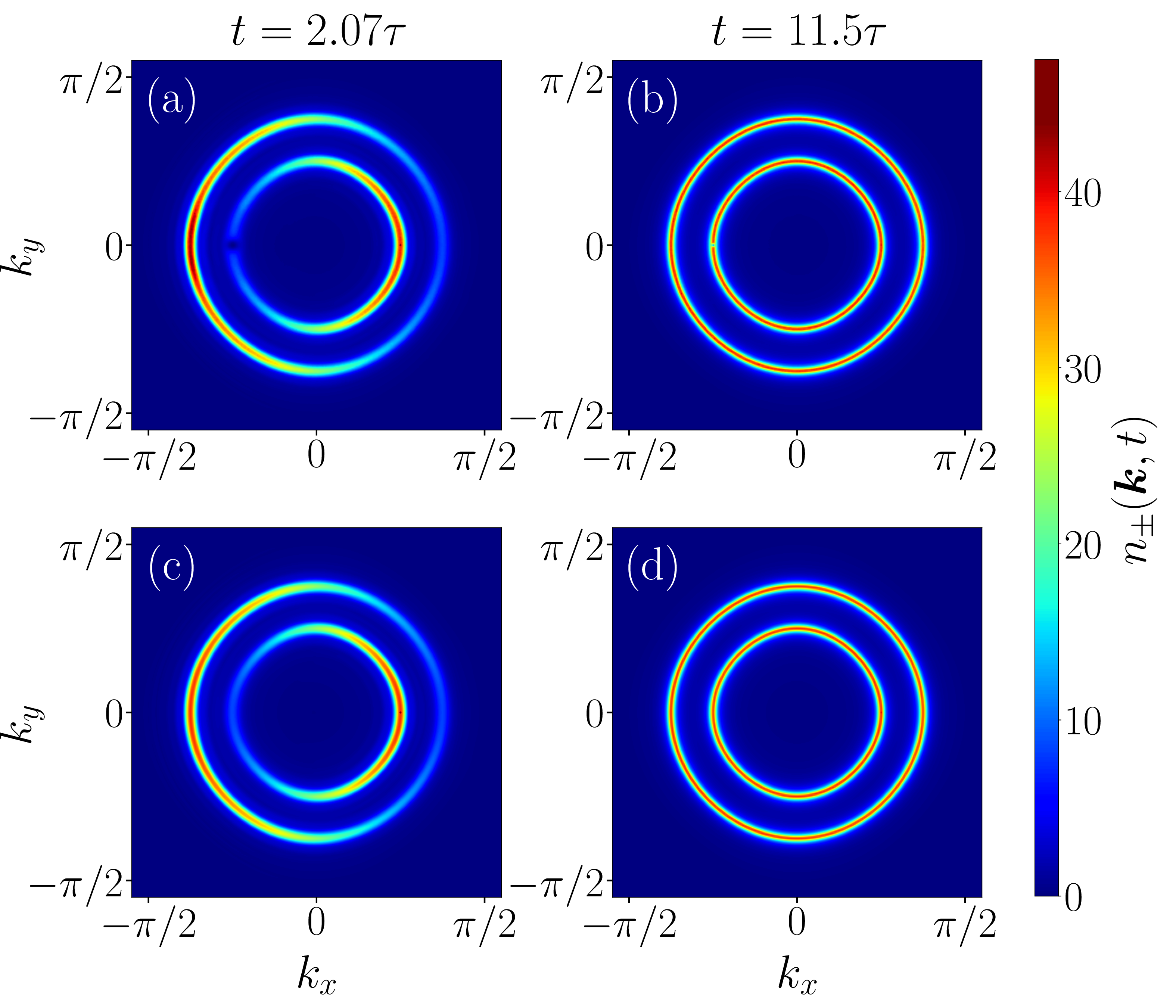}
  \caption{\label{fig:anisotropic_background}
  (Color online) The disorder-averaged momentum distribution in the strong SOC regime ($\varphi=\pi/16$) for $\bm{k}_0=(\pi/4,0)$.
  Simulation results at time (a)~$t=2.07\tau$ and (b)~$t=11.5\tau$, respectively. 
  Analytical results at time (c)~$t=2.07\tau$ and (d)~$t=11.5\tau$, respectively. 
  At $t=2.07\tau$, the disorder-averaged momentum distribution exhibits an anisotropic pattern, while at $t=11.5\tau$, it becomes isotropic except at $\bm{k}=-\bm{k}_0$. 
  }
\end{figure}
\begin{figure}[tbp]
  \includegraphics[width=0.8\linewidth]{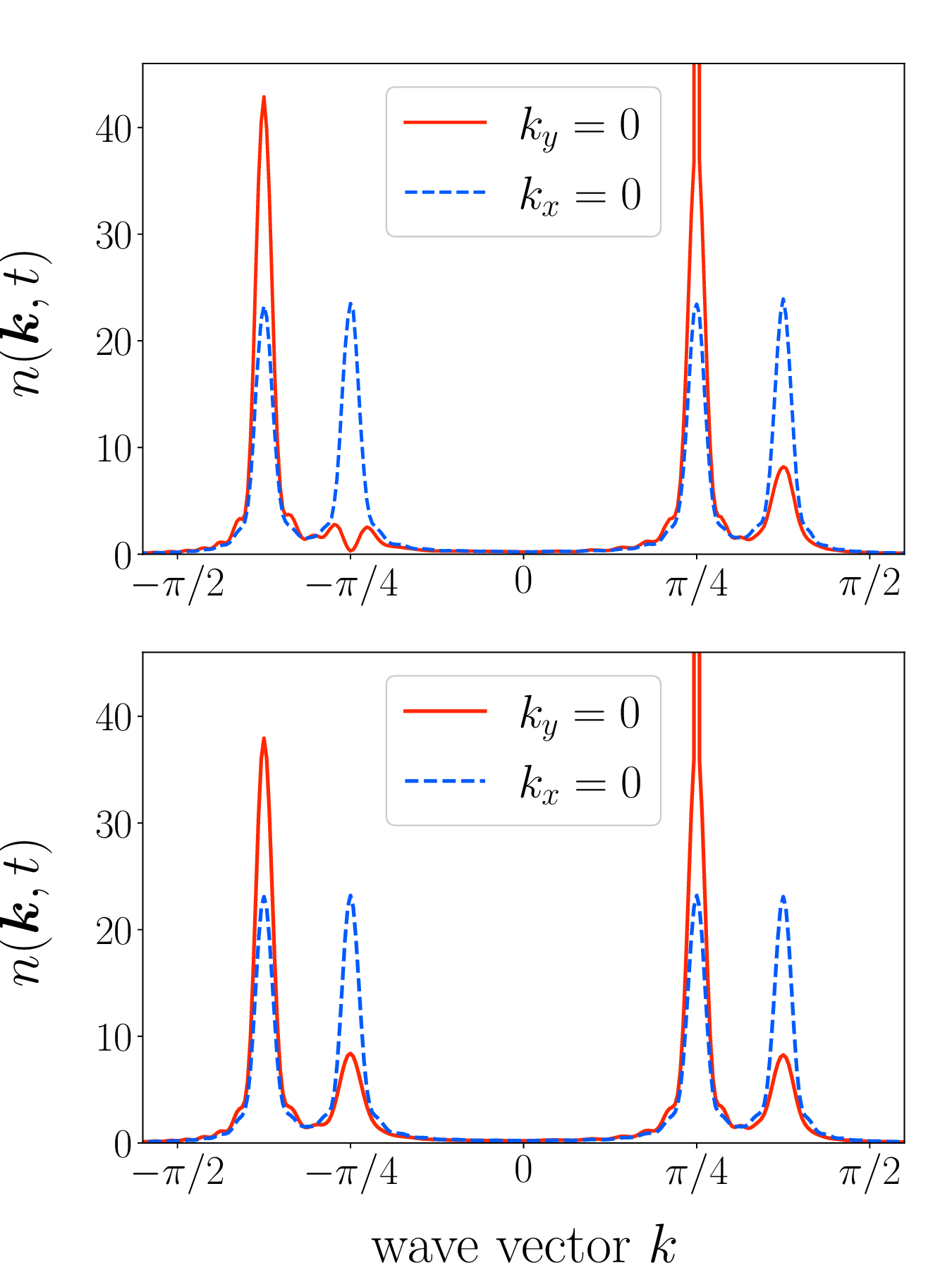}
  \caption{\label{fig:anisotropic_background_along_xy-axis}
  (Color online) 
  The disorder-averaged momentum distribution at time $t=2.07\tau$ plotted along the $k_x$ axis (red solid line) 
  and the $k_y$ axis (blue dashed line).
  (top)~Simulation results. (bottom)~Analytical results.
  These figures correspond to Fig.~\ref{fig:anisotropic_background}(a) and Fig.~\ref{fig:anisotropic_background}(c), respectively.
  The analytical calculation, considering only the Diffuson, is in agreement with the numerical simulation results, except at $\bm{k}=-\bm{k}_0$.
  }
\end{figure}
For strong SOC two diffusive rings are well resolved as a result of the spin splitting due to SOC.
These are clearly visible in Fig.~\ref{fig:3d_plot_strong} and also in Fig.~\ref{fig:anisotropic_background} and Fig.~\ref{fig:anisotropic_background_along_xy-axis}.
In Fig.~\ref{fig:anisotropic_background}, we compare the disorder-averaged momentum distribution obtained from simulations with those calculated using Eq.~(\ref{eq:analytic_n_other_direction}). 
In Figs.~\ref{fig:anisotropic_background}(a) and (b) we present the results of the simulations, while in Figs.~\ref{fig:anisotropic_background}(c) and (d) we present the results of the analytical calculations.
In Fig.~\ref{fig:anisotropic_background_along_xy-axis} we present cross-sectional views of Figs.~\ref{fig:anisotropic_background}(a) and (c) along the $k_x$ and $k_y$ axes, respectively.

\begin{figure}[tbp]
  \includegraphics[width=\linewidth]{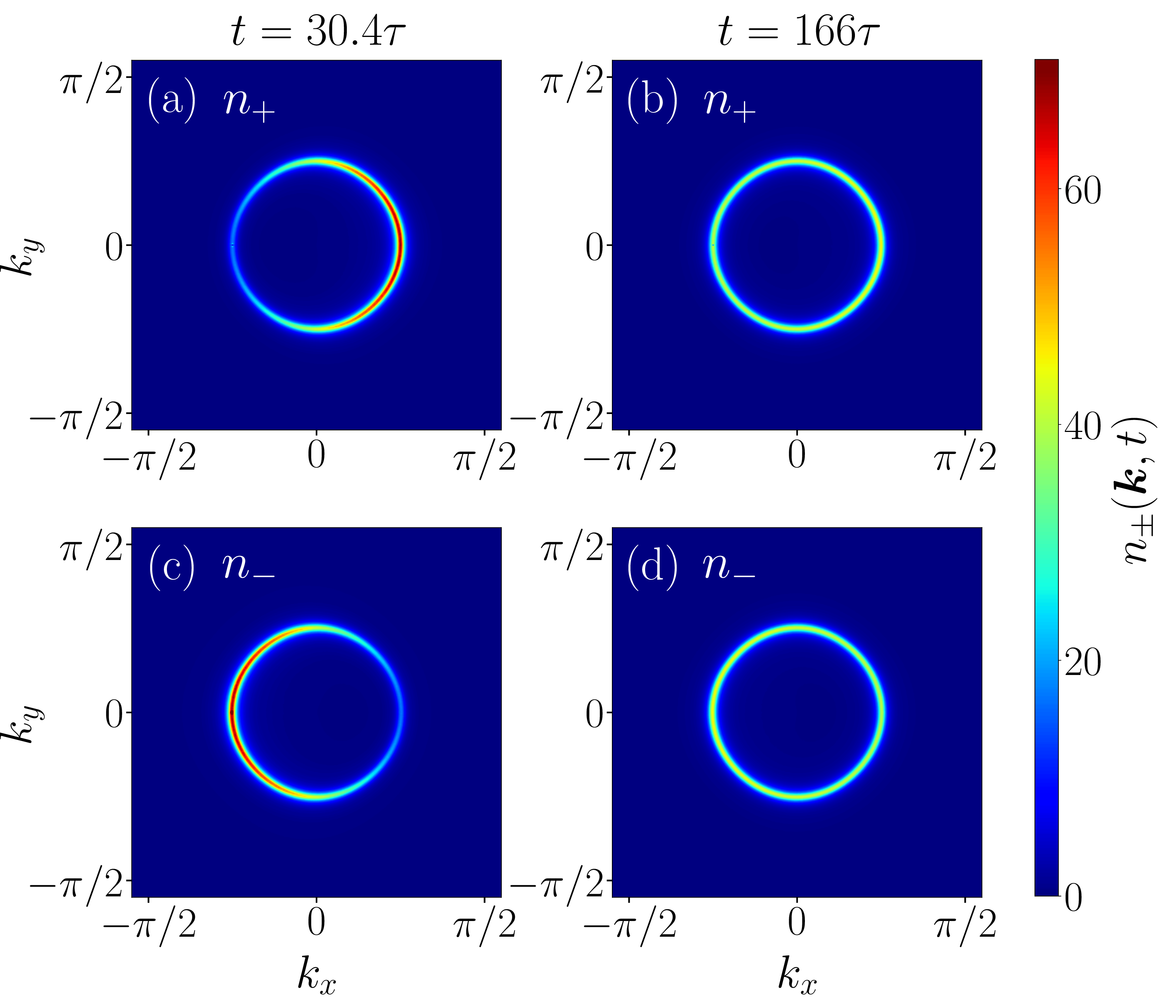}
  \caption{\label{fig:hidden-anisotropy_weak_SOC}
  (Color online) The disorder-averaged momentum distribution resolved into $+$ and $-$ components for weak SOC ($\varphi=\pi/1024$) for $\bm{k}_0=(\pi/4,0)$ obtained in simulations.
  The $+$ component $n_+(\bm{k})$ at time (a)~$t=30.4\tau$ and (b)~$t=166\tau$, respectively. 
  The $-$ component $n_-(\bm{k})$ at time (c)~$t=30.4\tau$ and (d)~$t=166\tau$, respectively. 
  The total distribution $n(\bm{k}) = n_-(\bm{k})+n_+(\bm{k})$ appears isotropic even at $t=30.4\tau$, even though each component is anisotropic.
  }
\end{figure}

For weak SOC the spin splitting is small, the two diffusive rings overlap,
and this anisotropy is hidden.
It is revealed, however, when the disorder-averaged momentum distribution is resolved into
$+$ and $-$ components as in Fig.~\ref{fig:hidden-anisotropy_weak_SOC}.

Comparing the results of the simulations with the analytical calculations, it is evident that the behaviour of the diffusive background can be explained using the Diffuson. 
The Cooperon contributes only in the vicinity of $\bm{k}=-\bm{k}_0$. 
The anisotropy in the disorder-averaged momentum distribution arises from the second term in Eq.~(\ref{eq:analytic_n_other_direction}). 
By examining the sign of this second term, we can deduce that intra-branch scattering is more likely to occur at small angles with respect to the initial wave vector than inter-branch scattering. 
Conversely, inter-branch scattering is more likely to occur at large angles. 
The second term in Eq.~(\ref{eq:analytic_n_other_direction}) decays exponentially in the long-time limit, as a result the imbalance between the components of the two branches vanishes in that limit. 
This corresponds precisely to the process of spin relaxation. 
The function $h$ given in Eq.~(\ref{eq:function_h}) decays at three different time scales, namely $\tau$, and
\begin{equation}
    \tau_1=\frac{1}{\omega_1},\ \ \tau_2=\frac{1}{\mathrm{Re}(\omega_2)}=\frac{1}{\mathrm{Re}(\omega_3)}.
\end{equation} 
Since the equilibration of the spin imbalance occurs on the slowest time scale, we conclude that
the spin relaxation time is given by $\tau_s=\tau_1$. 

\begin{figure}[tbp]
  \includegraphics[width=0.8\linewidth]{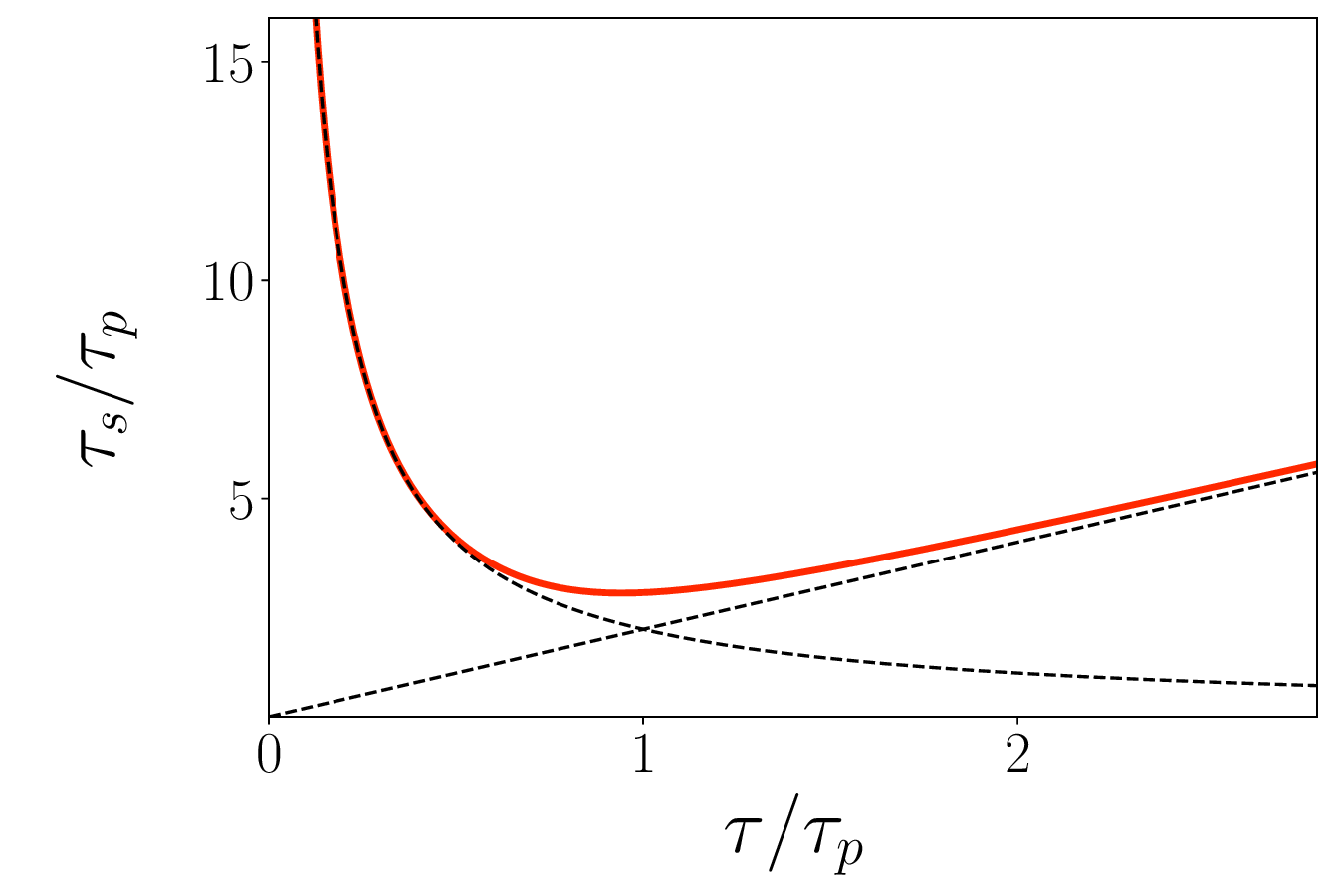}
  \caption{\label{fig:spin_relaxation_time}
  (Color online) The dependence of the spin relaxation time $\tau_s$ on the
  scattering time $\tau$.
  We make the abscissa and ordinate dimensionless by dividing them by the spin precession time $\tau_p = \hbar/\Delta$. 
  Note that $\tau_s/\tau_p$ is a function of a single variable, $\tau/\tau_p$. 
  The dashed lines show $2\tau_p/\tau$ and $2\tau/\tau_p$, respectively.
  }
\end{figure}

In Fig.~\ref{fig:spin_relaxation_time}, we show the dependence of the spin relaxation time $\tau_s$ on $\tau$. 
For weak SOC, the spin relaxation time is inversely proportional to the scattering time.
This behaviour is consistent with the D'yakonov-Perel' spin relaxation mechanism~\cite{Dyakonov_1972,Dyakonov_1986}. 
For strong SOC, we find Elliott-Yafet-like behaviour with the spin relaxation time proportional to $\tau$~\cite{Elliott_1954,Yafet_1963}.
The $\tau$ dependence of $\tau_s$ shown in Fig.~\ref{fig:spin_relaxation_time} provides a quantitative description
of the crossover between these two regimes of spin relaxation.

\section{Discussion}

We have investigated, both numerically and analytically, the time evolution of a particle in an initial plane wave 
state as it propagates coherently in a two-dimensional disordered system with SOC.
We have focused in particular on the time dependence of backscattering and the time dependence of the anisotropy of the diffusive background.
We have found that the results of our numerical simulations are well described by analytic calculations that treat the
SOC non-perturbatively and the effects of disorder using the Diffuson and Cooperon.

For weak SOC, we find at short times of the order of a few times the scattering time $\tau$
an enhancement of backscattering by a factor of two relative to the diffusive background.
This factor of two is characteristic of CBS in systems without SOC.
With SOC, however, for longer times, of the order of $(\tau_p / \tau)^2 \tau$, there is crossover to a reduction of 
backscattering by a factor of one half, which is characteristic of CBS in systems with SOC. 

With SOC, the spin polarisation of the initial plane wave state is reflected in an anisotropy of the disorder-averaged momentum distribution.
This anisotropy relaxes on the time scale of the spin relaxation time $\tau_s$ and we have taken advantage of this
to calculate the spin relaxation time.
For weak SOC, we have found a spin relaxation time that is inversely proportional to the scattering time $\tau$
consistent with the D'yakonov-Perel' spin relaxation mechanism.
For strong SOC, we have found an Elliott-Yafet-like behaviour with the spin relaxation time proportional to $\tau$.
Our analytic calculations describe quantitatively the crossover between these two regimes.
Our results are are consistent with those reported in Ref.~\cite{Szolnoki_2017_srep}.

The effects we discuss theoretically in this paper should be observable in a cold atomic gas with synthetic SOC.
For simplicity we have studied a model from solid state physics with an uncorrelated potential.
For easier comparison with experiments with cold atoms it may be important in future work to extend our calculations by using
a speckle potential.
It should be noted though that correlations of a realistic speckle potential impose a considerable burden in numerical simulations~\cite{Delande_2014,Orso_2017}. 
While taking account of these correlations is necessary, for example, for an accurate quantitative determination of the mobility
edge at which the Anderson transition occurs, we think the calculations we report do yield considerable insight.
Also, in systems with correlated disorder, such as a speckle potential, an anisotropy of the diffusive background emerges naturally~\cite{Cherroret_2012_PRA, Richard_2019}. 
However, in this study, the random potential is uncorrelated, and the anisotropy is purely a result of the SOC. 

\begin{figure}[t]
  \includegraphics[width=0.8\linewidth]{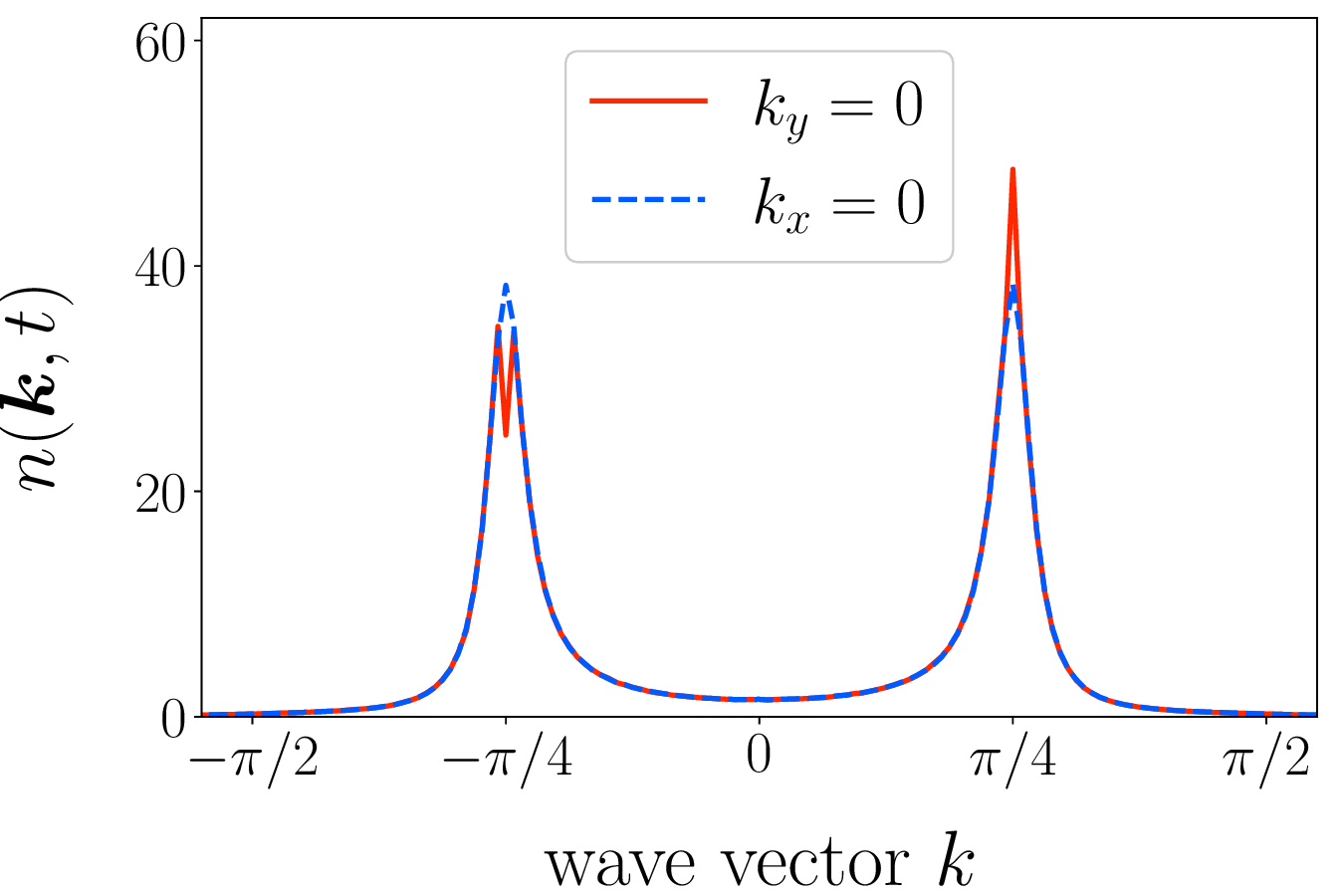}
  \caption{\label{fig:CFS}
  The disorder-averaged momentum distribution at time $t=1492\tau$ estimated by sampling 
  $8192$ disorder realisations for $\bm{k}_0=(\pi/4,0)$.
  The time $t$ is sufficiently longer than the Thouless time $\tau_D\simeq648\tau$ that a CFS peak is observed.}
\end{figure}

In the Anderson localised phase a coherent forward scattering (CFS) peak emerges in the disorder-averaged momentum distribution at $\bm{k}=\bm{k}_0$ at sufficiently long times~\cite{Karpiuk_2012,Arabahmadi_2024}.
Even in the metallic regime or in regimes where the localisation length is sufficiently long compared to the system size $L$, a CFS peak also manifests as an effect of finite size~\cite{Ghosh_2014}.
The relevant time scale is the Thouless time.
For the Ando model with $L=512$, $W=1$ and $\varphi=\pi/1024$, 
the estimated Thouless time is approximately $\tau_D\simeq522\tau$.
Therefore, no CFS peak is expected or observed in the simulations we reported above.
When simulating smaller systems and longer times a CFS peak appears.
In Fig.~\ref{fig:CFS} we show the CFS peak observed in a simulation of the Ando model with $L=256$, $W=1.5$, and $\varphi=\pi/1024$.
For these parameters, using the coherent potential approximation, we estimate $\tau=10.03$ in units of $\hbar/t_{\mathrm{hop}}$.

Finally, we have considered the non-interacting limit. The effects of interactions on coherent propagation in cold atoms have also been studied~\cite{Hartung_2008,Chretien_2021,Scoquart_2020} using the Gross-Pitaevskii equation.

\begin{acknowledgements}
    Keith Slevin is grateful for support from the Japan Society for the Promotion of Science under Grant-in-Aid 19H00658.
    Keith Slevin thanks Stefan Kettemann, Gabriel Lemari\'e, and Christian Minatura for helpful discussions.
\end{acknowledgements}

\appendix

\section{\label{app:matrix_element}Probability amplitude for scattering into a time reversed state}

We recall some basic definitions and properties based on Ref.~\cite{Haake_2007_book}.
For any two states $\left|\phi\right\rangle$ and $\left|\psi\right\rangle$ we have
\begin{equation}
    \left\langle \phi \right| \left. \psi \right\rangle = \left\langle \psi \right| \left. \phi \right\rangle^*.
\end{equation}
The hermitian conjugate of an operator $A$ is the operator $A^{\dagger}$ such that
\begin{equation}
    \left\langle \phi \right| \left. A \psi \right\rangle = \left\langle A^{\dagger} \phi \right| \left. \psi \right\rangle,
\end{equation}
for all states $\left|\phi\right\rangle$ and $\left|\psi\right\rangle$ satisfying the appropriate boundary conditions.
Time reversal operators are anti-unitary. The defining property of an anti-unitary operator is
\begin{equation}
   \left\langle T \phi \right| \left. T \psi \right\rangle = \left\langle \psi \right| \left. \phi \right\rangle.
\end{equation}
The time evolution operator for the system described by a time independent Hamiltonian $H$ is
\begin{equation}
  U\left(t\right) = \exp \left( -i H t \right),
\end{equation}
where $t$ is time.
Since any time reversal operator can be expressed as product of a unitary operator and complex conjugation in
a suitable basis, and assuming that 
\begin{equation}
    \left[ H, T \right] = 0,
\end{equation}
we see that
\begin{equation} \label{eq:TU}
    T U\left(t\right) = U^{\dagger}\left(t\right) T.
\end{equation}
We now consider the matrix element 
\begin{equation}
    \left\langle T \psi \right| \left. U\left(t\right) \psi \right\rangle.
\end{equation}
Since $T^2=-1$, this is equal to
\begin{equation}
    -\left\langle T \psi \right| \left. T^2 U\left(t\right) \psi \right\rangle.
\end{equation}
Using Eq.~(\ref{eq:TU}) this is then equal to
\begin{equation}
    - \left\langle T \psi\right| \left. T U^{\dagger}\left(t\right) T \psi \right\rangle.
\end{equation}
Since $T$ is anti-unitary, this is equal to
\begin{equation}
    -\left\langle U^{\dagger}\left(t\right) T \psi \right| \left. \psi \right\rangle.
\end{equation}
Using the definition of the hermitian conjugate, this is equal to
\begin{equation}
    -\left\langle T \psi \right| U\left(t\right) \left. \psi \right\rangle.
\end{equation}
Thus, we have found that
\begin{equation}
    \left\langle T \psi \right| \left. U\left(t\right) \psi \right\rangle = -\left\langle T \psi \right| \left. U\left(t\right) \psi \right\rangle 
\end{equation}
So the probability amplitude must be zero.
An alternative derivation that makes explicit use of Kramer's degeneracy is given in Ref.~\cite{Arabahmadi_2024}.

\section{\label{app:Green's func}Disorder-averaged Green's function of the Ando model}

The retarded free Green's function of the Ando model in the $\pm$~basis can be calculated using Eq.~(\ref{eq:eigenenergy_Ando}),
\begin{align}
    \nonumber
    G_0^R(\bm{k},\varepsilon) &= g_{0+}^R(\bm{k},\varepsilon)|\bm{k},+\rangle\langle\bm{k},+|\\ &\quad +g_{0-}^R(\bm{k},\varepsilon)|\bm{k},-\rangle\langle\bm{k},-|,
\end{align}
where $g_{0\pm}^R$ is
\begin{equation}
    g_{0\pm}^R(\bm{k},\varepsilon) = \frac{1}{\varepsilon-E_{\pm}(\bm{k})+i\delta}.
\end{equation}
Taking into account the relationship between the spin basis and the $\pm$~basis given by Eqs.~(\ref{eq:band_plus_state}) and (\ref{eq:band_minus_state}), the free retarded Green's function in the spin basis is expressed as
\begin{widetext}
\begin{equation}
    \mathrm{G}_0^R(\bm{k},\varepsilon) = \frac12\left(
    \begin{array}{cc}
        g_{0+}^R(\bm{k}, \varepsilon)+g_{0-}^R(\bm{k}, \varepsilon) & -e^{-i\theta(\bm{k})}\left(g_{0+}^R(\bm{k}, \varepsilon)-g_{0-}^R(\bm{k}, \varepsilon)\right)\\[1mm]
        -e^{i\theta(\bm{k})}\left(g_{0+}^R(\bm{k}, \varepsilon)-g_{0-}^R(\bm{k}, \varepsilon)\right) & g_{0+}^R(\bm{k}, \varepsilon)+g_{0-}^R(\bm{k}, \varepsilon)
    \end{array}
    \right).
\end{equation}
\end{widetext}
Within the scope of the Born approximation, the self-energy in the spin basis is 
\begin{align}\label{eq:self-energy_Born}
    \displaybreak[0]
    \nonumber
    \Sigma^R(\bm{k},\bm{k}',\varepsilon) &= \frac{1}{L^2}\sum_{\bm{k}''}
    \overline{
    \mathrm{V}(\bm{k},\bm{k}'')\mathrm{G}_0^R(\bm{k}'',\varepsilon)\mathrm{V}(\bm{k}'',\bm{k}')
    }\\
    &= \frac{1}{L^2}\sum_{\bm{k}''}
    \overline{
    \mathrm{V}(\bm{k},\bm{k}'')\mathrm{V}(\bm{k}'',\bm{k}')
    }\mathrm{G}_0^R(\bm{k}'', \varepsilon),
\end{align}
where $\mathrm{V}(\bm{k},\bm{k}')$ is $2\times2$ scalar matrix with
\begin{equation}
    \mathrm{V}_{\sigma\sigma'}(\bm{k},\bm{k}') = V(\bm{k}-\bm{k}')\delta_{\sigma,\sigma'},
\end{equation}
as elements.
Taking the disorder average, we obtain
\begin{equation}
    \overline{
    \mathrm{V}(\bm{k},\bm{k}'')\mathrm{V}(\bm{k}'',\bm{k}')
    } = \gamma(\bm{k}-\bm{k}'')\delta_{\bm{k},\bm{k}'}\mathrm{I}_2.
\end{equation}
Assuming that the random potential exhibits short-range correlations, we can take $\gamma(\bm{k})$ as a constant $\gamma_0$. 
Since $g_{0\pm}^{R}(-\bm{k},\varepsilon)=g_{0\pm}^{R}(\bm{k},\varepsilon)$ and $e^{i\theta(-\bm{k})}=-e^{i\theta(\bm{k})}$, when we perform the integration in Eq.~(\ref{eq:self-energy_Born}), the off-diagonal elements cancel out, and we
find that the self-energy $\Sigma^R(\bm{k},\varepsilon)$ is a scalar matrix with imaginary part
\begin{align}
    \nonumber
    &\mathrm{Im}\left\{\Sigma^R(\bm{k},\varepsilon)\right\}\\
    \nonumber
    \displaybreak[0]
    &\quad = \frac{\gamma_0}{L^2}\sum_{\bm{k}''}
    \frac{\mathrm{Im}\left\{g_{0+}^{R}(\bm{k}'', \varepsilon)\right\} + \mathrm{Im}\left\{g_{0-}^{R}(\bm{k}'', \varepsilon)\right\}}{2}\mathrm{I}_2\\
    \displaybreak[0]
    &\quad= -\frac{\pi}{2}\gamma_0\rho(\varepsilon)\mathrm{I}_2
\end{align}
Here, $\rho(\varepsilon)$ is the density of states per unit area
\begin{align}
    \nonumber
    \displaybreak[0]
    \rho(\varepsilon) &= \frac{1}{L^2}\sum_{\bm{k}''}
    \mathrm{Tr}\left[-\frac{1}{\pi}\mathrm{Im}\left\{\mathrm{G}_0^R(\bm{k}'',\varepsilon)\right\}\right]\\
    &= -\frac{1}{\pi L^2}\sum_{\bm{k}''}
    \Big[\mathrm{Im}\left\{g_{0+}^{R}(\bm{k}'', \varepsilon)\right\} + \mathrm{Im}\left\{g_{0-}^{R}(\bm{k}'', \varepsilon)\right\}\Big].
\end{align}
The scattering time is defined as the reciprocal of the imaginary part of the self-energy, thus
\begin{equation}
    \mathrm{Im}\left\{\Sigma^R(\bm{k},\varepsilon)\right\} = \frac{1}{2\tau}\mathrm{I}_2
\end{equation}
Since $\Sigma^R(\bm{k},\varepsilon)$ is a scalar matrix, it is unchanged when transforming to the $\pm$~basis.
As a result, the disorder-averaged Green's function in the $\pm$~basis is also given by Eq.~(\ref{eq:disorder-averaged_Greens_func_band_basis}).

\section{\label{app:approx_Gamma}Derivation of approximations of $\Gamma_1$, $\Gamma_2$ and $\Gamma_3$}

We define $E(\bm{k})$ and $\Delta(\bm{k})$ from the two energy eigenvalues in the Ando model,
\begin{align}
    \nonumber
    E_{\pm}(\bm{k}) &= -2t_1(\cos k_x + \cos k_y) \\
    \nonumber
    \displaybreak[0]
    &\quad\pm 2t_2\sqrt{\sin^2k_x + \sin^2k_y},\\
    &:= E(\bm{k})\pm\frac{\Delta(\bm{k})}{2}
\end{align}
In the integration we approximate $\Delta,\ \nu,\ \tau$ as constants, then using the residue theorem we obtain
\begin{align}
    \nonumber
    &\int\frac{d^2\bm{k}''}{(2\pi)^2}\,g_+^R\left(\bm{k}'',\varepsilon+\frac{\omega}{2}\right)g_+^A\left(\bm{k}'',\varepsilon-\frac{\omega}{2}\right)\\
    \nonumber
    \displaybreak[0]
    &\simeq \int\frac{\nu dE}{\left(\varepsilon-E-\frac{\Delta}{2}+\frac{\omega}{2}+\frac{i}{2\tau}\right)\left(\varepsilon-E-\frac{\Delta}{2}-\frac{\omega}{2}-\frac{i}{2\tau}\right)}\\
    \displaybreak[0]
    \label{eq:appendix_int1}
    &= \frac{2\pi i\nu}{\omega+\frac{i}{\tau}},\\
    \nonumber
    &\int\frac{d^2\bm{k}''}{(2\pi)^2}\,g_-^R\left(\bm{k}'',\varepsilon+\frac{\omega}{2}\right)g_-^A\left(\bm{k}'',\varepsilon-\frac{\omega}{2}\right)\\
    \displaybreak[0]
    &\simeq \frac{2\pi i\nu}{\omega+\frac{i}{\tau}},\\
    \nonumber
    &\int\frac{d^2\bm{k}''}{(2\pi)^2}\,g_+^R\left(\bm{k}'',\varepsilon+\frac{\omega}{2}\right)g_-^A\left(\bm{k}'',\varepsilon-\frac{\omega}{2}\right)\\
    \displaybreak[0]
    &\simeq \frac{2\pi i\nu}{\omega+\Delta+\frac{i}{\tau}},\\
    \nonumber
    &\int\frac{d^2\bm{k}''}{(2\pi)^2}\,g_-^R\left(\bm{k}'',\varepsilon+\frac{\omega}{2}\right)g_+^A\left(\bm{k}'',\varepsilon-\frac{\omega}{2}\right)\\
    \label{eq:appendix_int4}
    &\simeq \frac{2\pi i\nu}{\omega-\Delta+\frac{i}{\tau}}.
\end{align}
Using Eqs.~(\ref{eq:appendix_int1})-(\ref{eq:appendix_int4}), we can calculate approximations for $\Pi_1$ and $\Pi_2$ from Eqs.~(\ref{eq:def_Pi1}) and (\ref{eq:def_Pi2}). Substituting them into Eqs.~(\ref{eq:def_Gamma1})-(\ref{eq:def_Gamma3}), their denominators are
\begin{align}
    \nonumber
    &1-\gamma_0\left(\Pi_0+\Pi_1\right)\\
    \nonumber
    &= 1-\frac12\int\frac{d^2\bm{k}''}{(2\pi)^2}\left[g_+^R\left(\bm{k}'',\varepsilon+\frac{\omega}{2}\right)g_+^A\left(\bm{k}'',\varepsilon-\frac{\omega}{2}\right)\right.\\ 
    \nonumber
    &\hspace{80pt}\left.+ g_-^R\left(\bm{k}'',\varepsilon+\frac{\omega}{2}\right)g_-^A\left(\bm{k}'',\varepsilon-\frac{\omega}{2}\right)\right]\\
    \displaybreak[0]
    &\simeq 1 - \frac{\frac{i}{\tau}}{\omega+\frac{i}{\tau}} = \frac{\omega}{\omega+\frac{i}{\tau}},\\[1.5mm]
    \nonumber
    &1-\gamma_0\Pi_0\\
    \nonumber
    &\simeq \frac14\left[\frac{\omega+\Delta}{\omega+\Delta+\frac{i}{\tau}} + \frac{\omega-\Delta}{\omega-\Delta+\frac{i}{\tau}} + \frac{2\omega}{\omega+\frac{i}{\tau}}\right]\\
    \displaybreak[0]
    &= \frac{\left(\omega+i\omega_1\right)\left(\omega+i\omega_2\right)\left(\omega+i\omega_3\right)}{\left(\omega+\frac{i}{\tau}\right)\left(\omega+\Delta+\frac{i}{\tau}\right)\left(\omega-\Delta+\frac{i}{\tau}\right)},\\[1.5mm]
    \nonumber
    &1-\gamma_0\left(\Pi_0-\Pi_1\right)\\
    \nonumber
    &\simeq \frac12\left[\frac{\omega+\Delta}{\omega+\Delta+\frac{i}{\tau}} + \frac{\omega-\Delta}{\omega-\Delta+\frac{i}{\tau}}\right]\\
    &= \frac{\left(\omega+i\omega_4\right)\left(\omega+i\omega_5\right)}{\left(\omega+\Delta+\frac{i}{\tau}\right)\left(\omega-\Delta+\frac{i}{\tau}\right)}.
\end{align}
Thus, we obtain Eqs.~(\ref{eq:approx_Gamma_1})-(\ref{eq:approx_Gamma_3}).

\newpage

\bibliography{refs}

\begin{thebibliography}{58}
\bibitem [{Roa()}]{Roati_note}%
  \BibitemOpen
  \href@noop {} {}\bibinfo {note} {Simultaneously, the following paper has been published demonstrating Anderson localisation using a quasi-periodic optical lattice}\BibitemShut {NoStop}%
\bibitem [{Tim()}]{Time-reversal_note}%
  \BibitemOpen
  \href@noop {} {}\bibinfo {note} {We note that the phase factors in Eqs.~(\ref{eq:TR-state1}) and (\ref{eq:TR-state2}) are to some extent arbitray, as the eigenvectors~(\ref{eq:band_plus_state}) and (\ref{eq:band_minus_state}) and the time reversal operator $T$ are only defined up to arbitary complex phase factors. These phase factors do not affect the subsequent calculations.}\BibitemShut {Stop}%
\bibitem [{int()}]{integral_note}%
  \BibitemOpen
  \href@noop {} {}\bibinfo {note} {In Sec.~\ref{sec:diagrammatic_calculation}, all sums with respect to $\bm{k}''$ are replaced by integrals.}\BibitemShut {Stop}%
\end{thebibliography}

\end{document}